\documentclass[]{aa}

\usepackage[varg]{txfonts}

\usepackage{lineno}
\usepackage{natbib}
\usepackage{amsmath}
\usepackage{graphicx}
\usepackage{wasysym}
\usepackage{txfonts}
\usepackage{xspace}
\usepackage{url}
\usepackage{textcomp}
\usepackage{multirow}
\usepackage{lipsum}

\newcommand{\gx}{GX~301$-$2\xspace}

\newcommand{\swift}{\textsl{Swift}\xspace}
\newcommand{\fermi}{\textsl{Fermi}\xspace}

\newcommand{\inte}{\textsl{INTEGRAL}\xspace}

\newcommand{\xmm}{\textsl{XMM-Newton}\xspace}
\newcommand{\sax}{\textsl{BeppoSAX}\xspace}
\newcommand{\suz}{\textsl{Suzaku}\xspace}

\newcommand{\xte}{\textsl{RXTE}\xspace}

\newcommand{\nustar}{\textsl{NuSTAR}\xspace}
\newcommand{\highe}{\texttt{highecut}\xspace}
\newcommand{\fdcut}{\texttt{FDcut}\xspace}
\newcommand{\npex}{\texttt{NPEX}\xspace}

\newcommand{\asec}{\ensuremath{''}\xspace}

\newcommand{\snr}{S/N\xspace}

\newcommand{\msun}{\ensuremath{\text{M}_{\odot}}\xspace}

\newcommand{\redchi}{\ensuremath{\chi^{2}_\text{red}}\xspace}

\newcommand{\feka}{\ensuremath{\mathrm{Fe}~\mathrm{K}\alpha}\xspace}
\newcommand{\fekb}{\ensuremath{\mathrm{Fe}~\mathrm{K}\beta}\xspace}
\newcommand{\nika}{\ensuremath{\mathrm{Ni}~\mathrm{K}\alpha}\xspace}

\newcommand{\nh}{\ensuremath{{N}_\mathrm{H}}\xspace}

\newcommand{\nhtwo}{\ensuremath{{N}_{\mathrm{H},2}}\xspace}

\newcommand{\ergcms}{\ensuremath{\text{erg\,cm}^{-2}\text{s}^{-1}}}
\newcommand{\ergs}{\ensuremath{\text{erg}\,\text{s}^{-1}}}

\begin{document}

\title{Multiple cyclotron line-forming regions in  \gx}

\author{F.~F\"urst\inst{1}\and S.~Falkner\inst{2}\and D.~Marcu-Cheatham\inst{3,4}\and B.~Grefenstette\inst{5}\and J.~Tomsick\inst{6}\and K.~Pottschmidt\inst{3,4} \and D.~J.~Walton\inst{7} \and L.~Natalucci\inst{8} \and P.~Kretschmar\inst{1}}

\institute{European Space Astronomy Centre (ESAC), Science Operations Department, 28692 Villanueva de la Ca\~nada, Madrid, Spain
\and Dr. Karl-Remeis-Sternwarte and ECAP, Sternwartstr. 7, 96049 Bamberg, Germany
\and CRESST, Department of Physics, and Center for Space Science and Technology, UMBC, Baltimore, MD 21250, USA
\and NASA Goddard Space Flight Center, Greenbelt, MD 20771, USA
\and Cahill Center for Astronomy and Astrophysics, California Institute of Technology, Pasadena, CA 91125, USA
\and Space Sciences Laboratory, University of California, Berkeley, CA 94720, USA
\and Institute of Astronomy, Madingley Road, Cambridge CB3 0HA, UK
\and INAF-Institute for Space Astrophysics and Planetology, Via Fosso del Cavaliere 100, 00133 Rome, Italy
}

\abstract{
We present two observations of the high-mass X-ray binary \gx with \nustar, taken at different orbital phases and different luminosities. We find that the continuum is well described by typical phenomenological models, like a very strongly absorbed NPEX model. However, for a statistically acceptable description of the hard X-ray spectrum we require two cyclotron resonant scattering features (CRSF), one at $\sim$35\,keV and the other at $\sim$50\,keV. Even though both features strongly overlap, the good resolution and sensitivity of \nustar allows us to disentangle them at $\ge99.9\%$ significance. This is the first time that two CRSFs are seen in \gx. We find that the CRSFs are very likely independently formed, as their energies are not harmonically related and, if it were a single line,  the deviation from a Gaussian shape would be very large.
We compare our results to archival \suz data and find that our model also provides a good fit to those data.
We study the behavior of the continuum as well as the CRSF parameters as function of pulse phase in seven phase bins. 
We find that the energy of the 35\,keV CRSF varies smoothly as function of phase, between 30--38\,keV. To explain this variation, we apply a simple model of the accretion column, taking the altitude of the line-forming region, the velocity of the in-falling material, and the resulting relativistic effects into account. We find that in this model the observed energy variation can be explained  simply due to a variation of the projected velocity and beaming factor of the line forming region towards us.
} 


\date{Received XX.XX.XX / Accepted XX.XX.XX}

\maketitle

\section{Introduction}
\label{sec:intro}

Cyclotron resonant scattering features (CRSFs, or cyclotron lines) in the hard X-ray spectra of accreting neutrons stars allow for the measurement of the magnetic field strength close to the neutron star surface. CRSFs have been detected in $\sim$25 sources to date \citep{caballero12a} and the measurements show that these neutron stars possess  strong magnetic fields of the order of $10^{12}$\,G \citep[e.g.,][]{truemper78a}. These strong fields dominate the accretion geometry, leading to the formation of accretion columns above the magnetic poles in which most of the hard X-ray radiation is produced. To understand the physics within these accretion columns, as well as the emission geometry of the X-rays, a thorough understanding of the magnetic field configuration is therefore important.  

CRSFs show up as broad absorption features in the hard X-ray spectrum of accreting pulsars. They are formed by scattering of photons off electrons quantized onto Landau-levels in the strong magnetic field close to the surface of the neutron star. As the Landau levels are directly correlated with the magnetic field strength, so is the CRSF energy, and we can estimate the magnetic field strength in the line-forming region following the so-called 12-B-12 rule,
\begin{equation}
B_{12} = E_\text{CRSF} \times (1+z)  / 11.6
\end{equation}
Here, $B_{12}$ is the magnetic field in $10^{12}$\,G, $E_\text{CRSF}$ is the cyclotron line energy in keV, and $z$ is the gravitational redshift due to the mass of the neutron star.

In recent years, more and more evidence has been collected that the CRSF energy is variable as function of luminosity \citep[see, e.g.,][and references therein]{velax1nustar}. The direction of the correlation seems to depend on luminosity: at the very high luminosities ($L\gtrsim 3\times10^{37}\,\ergs$), an anti-correlation between CRSF energy and luminosity is evident \citep[e.g., V\,0332+53;][]{tsygankov10a}, while at lower luminosities a positive correlation has been found \citep[e.g., Her~X-1;][]{staubert07a}. Different theories for this behavior have been proposed, e.g., the formation of a shock in the accretion column \citep{becker12a}, or reflection on the neutron star surface \citep{poutanen13a}.

At the very lowest luminosities, below $\sim5\times10^{36}$\,\ergs, the situation gets even more complicated: sources located there do not show a uniform behavior. For example, the CRSF energy of A\,0535+26 seems to be constant as a function of luminosity \citep{caballero07a, ballhausen17a}, while in Vela~X-1 a clear positive trend is observed \citep{velax1nustar}. Theoretical work explaining the different behavior at these luminosities is currently still missing.

While the CRSFs give us a good idea about the magnetic field strength, measurements of the magnetic field geometry are not straight forward.
To first order, the magnetic field can be assumed to be a  simple symmetric dipole. This approach explains, for example, the apparent variation of the CRSF energy with pulse phase in \gx, assuming that at different phases different magnetic field strengths are sampled \citep{suchy12a}. However, it fails to explain the dependence of the CRSF energy in Cen~X-3 \citep{suchy08a}, for which a dipole magnetic field, off-set from the neutron star center, might provide a better explanation \citep{kraus96a}. 

Decomposition of pulse-profiles also often results in the requirement of an asymmetric dipole field \citep[e.g.,][and references therein]{sasaki12a}. In this approach, the energy dependent pulse profile is modeled with phase-dependent contributions from the two accretion poles, under the influence of strong light-bending of the emitted photons by the gravitational field of the neutron star. For a good description of the observed pulse profile this decomposition typically requires that the two poles are not exactly opposite of each other, but slightly offset \citep[e.g., $\sim25^\circ$ in A\,0535+26;][]{caballero11a}.

In a detailed study of 4U~0115+63, the source with the most CRSFs known \citep[up to 5;][]{santangelo99a, heindl99a}, \citet{iyer15a} find evidence that the lines can be separated into two pairs, with different fundamental energies. They argue that this separation could be explained if accretion happens on both poles,  but each pole has a different magnetic field strength, due to an off-set magnetic field from the neutron star center.

Even more complicated magnetic field geometries are sometimes proposed  for the  recently discovered ultra-luminous pulsars \citep[e.g.,][]{bachetti14a}. While the spin-up rate of NGC\,7793~P13 seems to agree with a dipolar field of $1.5\times10^{12}$\,G \citep{p13}, \citet{israel17a} need to invoke a strong multi-polar component to explain the extreme luminosity of NGC\,5907~ULX1.

\begin{figure}
\begin{center}
\includegraphics[width=0.95\columnwidth]{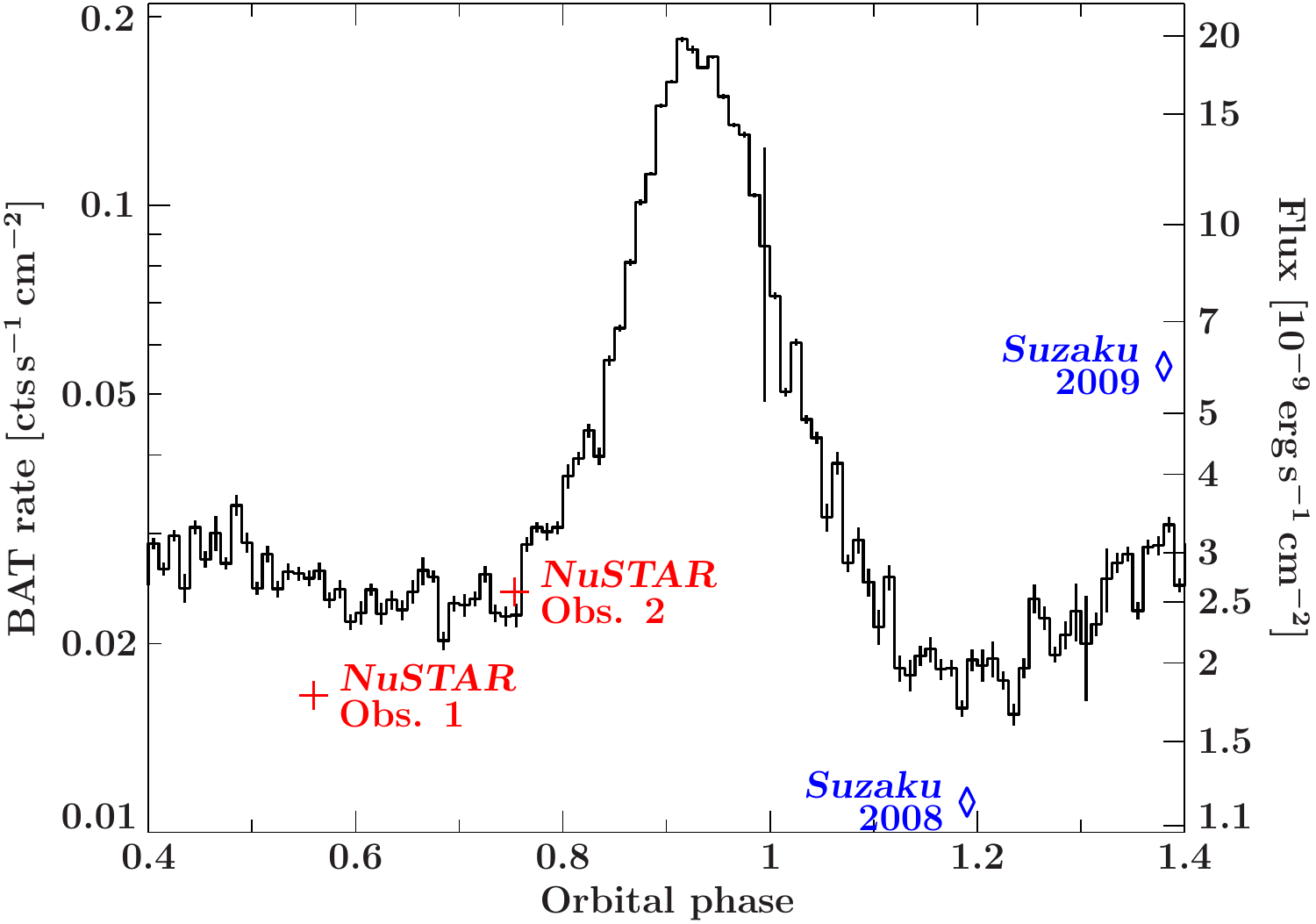}
\caption{Average \swift/BAT light-curve, folded on the orbital period where orbital phase 1 is periastron. A strong flare shortly before periastron is clearly visible.  Note that due to the eccentric orbit the phases are not equidistant in time. The two \nustar observations are shown by the red crosses and the \suz observations by the blue diamonds. These data use the right-hand $y$-scale with the flux measured between 5--50\,keV. The left and right $y$-scales are aligned based on the daily BAT count-rate measured during the \nustar observations. Note the strong variability above and below the average flux in the pointed observations of \nustar and \suz. }
\label{fig:batlc}
\end{center}
\end{figure}

To study the magnetic field geometry as well as the luminosity dependence of the CRSF energy at low luminosities, the high mass X-ray binary (HMXB) \gx is an ideal target. It has a high pulsed fraction, a well-studied strong CRSF and, due to it being only about 3\,kpc away, also has a high flux at low luminosities. It was discovered in 1969 by balloon-born experiments \citep{lewin71b, mcclintock71a}. Its pulse period is around $\sim$683\,s \citep{white76a} and the system has an orbital period of $\sim$41.5\,d \citep{koh97a, doroshenko10a} with a relatively large eccentricity of $e=0.47$. This eccentricity gives rise to a regular pre-periastron flare, at which the neutron star overtakes the accretion stream \citep{leahy02a, leahy08a}. Therefore, the average flux is strongly orbital-phase dependent, varying between 2--20$\times10^{-9}$\,\ergcms (Fig.~\ref{fig:batlc}). However, strong variability on top of that average luminosity is present, in particular outside the pre-periastron flare, where accretion is dominated by direct accretion from the stellar wind. As an example, Fig.~\ref{fig:batlc} shows the fluxes of two archival \suz observations, which differ strongly from the expected average flux at their respective orbital phases.

\citet{mihara95a} discovered a CRSF in the hard X-ray spectrum of \gx using \textsl{Ginga} data. They describe a very broad ($\sigma \approx 16$\,keV) and strong line around 35\,keV. The line was later confirmed by \citet{kreykenbohm04a}, using \xte data. These authors also found that the line energy is varying between $\sim$30--40\,keV as a function of pulse phase. The variation was studied in more detail by \citet{suchy12a}, who explain the smooth sinusoidal variation of the energy as function of pulse phase by different viewing angles on a dipolar magnetic field.
\citet{labarbera05a} analyzed four \sax observations, and also found a CRSF, albeit at much higher energies, at around 45--53\,keV, depending on orbital phase. These energies are consistent with the values reported by \citet{doroshenko10a}, who analyzed \inte data taken during the pre-periastron flare.
Note that in all studies, the CRSF is very strong and removes a significant fraction of the continuum flux. An accurate modeling therefore depends on obtaining the correct continuum shape, and also requires good coverage of the energies above and below the ones dominated by the CRSF.

The large width of the CRSF also resulted in relatively large uncertainties of the centroid energy of the line, making a detailed study of the behavior of the line energy as function of luminosity difficult. \citet{suchy12a} see an indication for an anti-correlation between energy and luminosity, however, the trend is not significant at the 90\% level. Combining the results by \citet{suchy12a} with the measurements by \citet{labarbera05a} a positive correlation seems to be present but it is very likely that these values are dominated by systematic differences in the analyses.

In this paper we analyze two observations of \gx taken with the \textsl{Nuclear Spectroscopic Telescope Array} \citep[\nustar,][]{harrison13a}. The \nustar data provide unprecedented energy resolution and signal-to-noise ratio (\snr) above 10\,keV and are therefore ideally suited to study the CRSF region.
In Sect.~\ref{sec:data} we describe the data reduction and calibration. In Sect.~\ref{sec:phasavg} we present phase-averaged spectroscopy and compare the spectrum to archival \suz data. In Sect.~\ref{sec:phasres} we discuss the pulse profiles and  perform phase-resolved spectroscopy. We discuss a physical interpretation of the results and draw conclusions in Sect.~\ref{sec:disc}.

\section{Data analysis \& calibration}
\label{sec:data}

The two \nustar observations are separated by roughly one year and  were performed at different orbital phases (see Table~\ref{tab:obslog}). The first observation was performed just after apastron, during an expected intermediate flux state of the source. However, by chance the observations took place at a much lower luminosity than expected. The second observation was obtained during a later orbital phase and closer to the pre-periastron flare, which resulted in the desired higher fluxes (Fig.~\ref{fig:batlc}). As far as we know, these data have not been published elsewhere yet.

\nustar has two identical instruments, Focal Plane Module A and B (FPMA and FPMB, respectively). We extract data from both modules separately, using the standard pipeline \texttt{nustardas} v1.7.1, as distributed with HEASOFT v6.20 and calibration files 20170222. Source data  from both observations were extracted from a circular region with a radius of 40\asec, centered on the brightest spot in the respective sky images of each module. The extraction region size was chosen to optimize the \snr at  energies above 45\,keV. A larger region introduces more background photons than source photons at the highest energies, due to the PSF shape. The influence of the smaller region onto the \snr at low energies is negligible.
Background data were extracted from a circular region with a radius of 140\asec, located at the opposite end of the field-of-view, to avoid contamination from source photons.
The spectra were rebinned following the energy resolution of the detectors, by combining  2,3,5,8,16,18,48, and 72 raw channels for energies above 3,10,15,20,35,45,55, and 65\,keV, respectively. Additionally we required at least 20 counts per bin and modeled the data in the 3--78\,keV energy range. 

\suz observed \gx twice, as described by \citet{suchy12a}. Here we reanalyze the longer of the two observations, ObsID 4030444020,  which was taken on 2009, January 05 (Tab.~\ref{tab:obslog}) and took place at post-periastron phase (0.38). The data were obtained with the X-ray Imaging Spectrometer \citep[XIS,][]{koyama07a} with CCD cameras 0, 1, 3 operating in normal mode in a 1/4 window sub-array option, along with data from the PIN instrument from the Hard X-ray Detector \citep[HXD,][]{takahashi07a}. The exposure times for each instrument were 61.8\,ks for XIS and 51.0\,ks for PIN (Tab.~\ref{tab:obslog}).

The \textsl{Suzaku} data were reprocessed and extracted using the standard \texttt{aepipeline} as distributed with HEASOFT v6.19 and calibration packages HXD-20110913, and XIS-20160607, and the screened XIS events were filtered to exclude times of telemetry saturation. The XIS images were first extracted with \texttt{xselect}, and then further corrected for thermal attitude uncertainties with \texttt{aeattcor2}. For both $3\times3$ and $5\times5$ editing modes, the source data were extracted from circular regions with $120''$ radii, with $20''$ pile-up exclusion radii centered on the PSFs. The background regions, circular with $120''$ radii, were extracted at the furthest locations from the point source within the windows.\footnote{We avoided the dead zone of XIS\,0 (Tsujimoto et al. 2010) during the background region selection.} We generated the XIS source and background event files, images, and spectra using \texttt{xselect}.
The energy and ancillary responses were generated with \texttt{xisrmfgen} and \texttt{xissimarfgen}. 

Using hxdpinxbpi, we extracted the dead-time corrected PIN source spectrum and the total background spectrum 
and applied the \verb+ae_hxd_pinhxnome5_20080716.rsp+ PIN response file for the specific calibration epoch. The total PIN background spectrum is composed of the sum of cosmic X-ray background (CXB; $\sim$5\% of the PIN background flux) and non X-ray modeled background (NXB; $\sim$95\% of the PIN background flux).

Data analysis was performed with the Interactive Spectral Interpretation System \citep[ISIS,][]{houck00a} v1.6.2. All uncertainties are reported at the 90\% confidence level unless otherwise noted.

\begin{table}
\caption{Observation log}
\begin{center}
\begin{tabular}{llll}
ObsID & Start date & Exposure & Orb. Phase \\\hline
\nustar \\
30001041002 & 2014-10-29 & 38.2\,ks  & 0.55 \\
30101042002 & 2015-10-04 & 35.7\,ks & 0.75 \\\hline
\suz \\
403044020 & 2009-01-05 & 61.8\,ks (XIS)  & 0.38 \\
 & & 51.0\,ks (PIN)  & \\\hline
\end{tabular}
\end{center}
\label{tab:obslog}
\end{table}%

\section{Phase-averaged spectroscopy}
\label{sec:phasavg}

\subsection{\nustar observation 2}

As the second observation (ObsID 30101042002) was the overall brighter observation with a higher \snr, we start our spectral modeling with these data and later apply our preferred models to the other \nustar observation, as well as to archival \suz data. The spectrum of \gx has been well studied in X-rays so we can already start with a relatively sophisticated spectral model, including a partially covering absorber (with column density \nh and covering fraction $f$). We model the absorption with an updated version of the \texttt{tbabs} model \citep{wilms00a}, using the \texttt{wilm} abundance vector and cross-sections presented by \citet{verner96a}.
We add \feka and \fekb fluorescence lines (with line center energy $E_{\text{K}\alpha/\beta}$, width $\sigma_{\text{K}\alpha/\beta}$, and equivalent width $\text{EW}_{\text{K}\alpha/\beta}$).
However, due to the proximity to the Fe K-edge, we cannot constrain the width of the \fekb line, and therefore fixed it to be narrow ($\sigma_{\text{K}\beta}=10^{-6}$\,keV).
We apply different phenomenological models to describe the continuum emission, as no physical model is yet available that describes this emission at the relatively low luminosity of \gx during our observations ($\sim2\times10^{36}\,\ergs$).

To model the previously observed CRSF in the hard X-ray spectrum, we use a multiplicative absorption model with a Gaussian optical depth profile (model \texttt{gabs} in XSPEC, described through its energy $E_\text{CRSF}$, width $\sigma_\text{CRSF}$, and strength $d_\text{CRSF}$\footnote{the strength $d$  is related to the optical depth $\tau$ at the line center as $\tau =  d/( \sigma \sqrt{2\pi})$}). From previous investigations at similar luminosities \citep[e.g.,][]{suchy12a} we expect the CRSF energy to be around 35\,keV, and we initialize the model with this energy.
We also add a multiplicative constant to the model for FPMB, to allow for slight differences in absolute flux calibration with respect to FPMA (CC$_\text{B}$).

This baseline model can thus be written as
\begin{equation}
\text{CC}\times\bigl(f\times\nh + \left[1-f\right]\bigr)\times\bigl(\text{CONT}\times\text{CRSF}_1 + \feka + \fekb \bigr)
\end{equation}
where CONT is the respective continuum model.

Consistent with the very high absorption column and prominent iron lines, \gx is known to also show a Compton shoulder at the red wing of the \feka line \citep{watanabe03a, gx301xmm}. However, this shoulder extends only to about 6.2\,keV and is within the energy resolution of \nustar, completely subsumed by the Gaussian model for the \feka line. We therefore do not include this shoulder in our spectral modeling. Small residuals at the red end of the line might be related to this shoulder, or to the unidentified absorption feature found around 5.6\,keV in \xmm data by  \citet{gx301xmm}. 

 We apply different phenomenological models typically used to describe the continuum of magnetized, accreting neutrons stars: \texttt{cutoffpl}, \fdcut \citep{tanaka86a}, \npex \citep{mihara98a}, and \texttt{HIGHECUT}. For the latter, we smoothed the artificial kink at the cutoff energy with an additional multiplicative Gaussian line.
The continuum parameters are, where applicable, the flux $\mathcal{F}$ between 5--50\,keV, the cutoff energy $E_\text{cut}$, the folding energy $E_\text{fold}$, the photon index $\Gamma$, and the normalization of the second component $A_2$ in the \npex model. 

\begin{figure}
\begin{center}
\includegraphics[width=0.95\columnwidth]{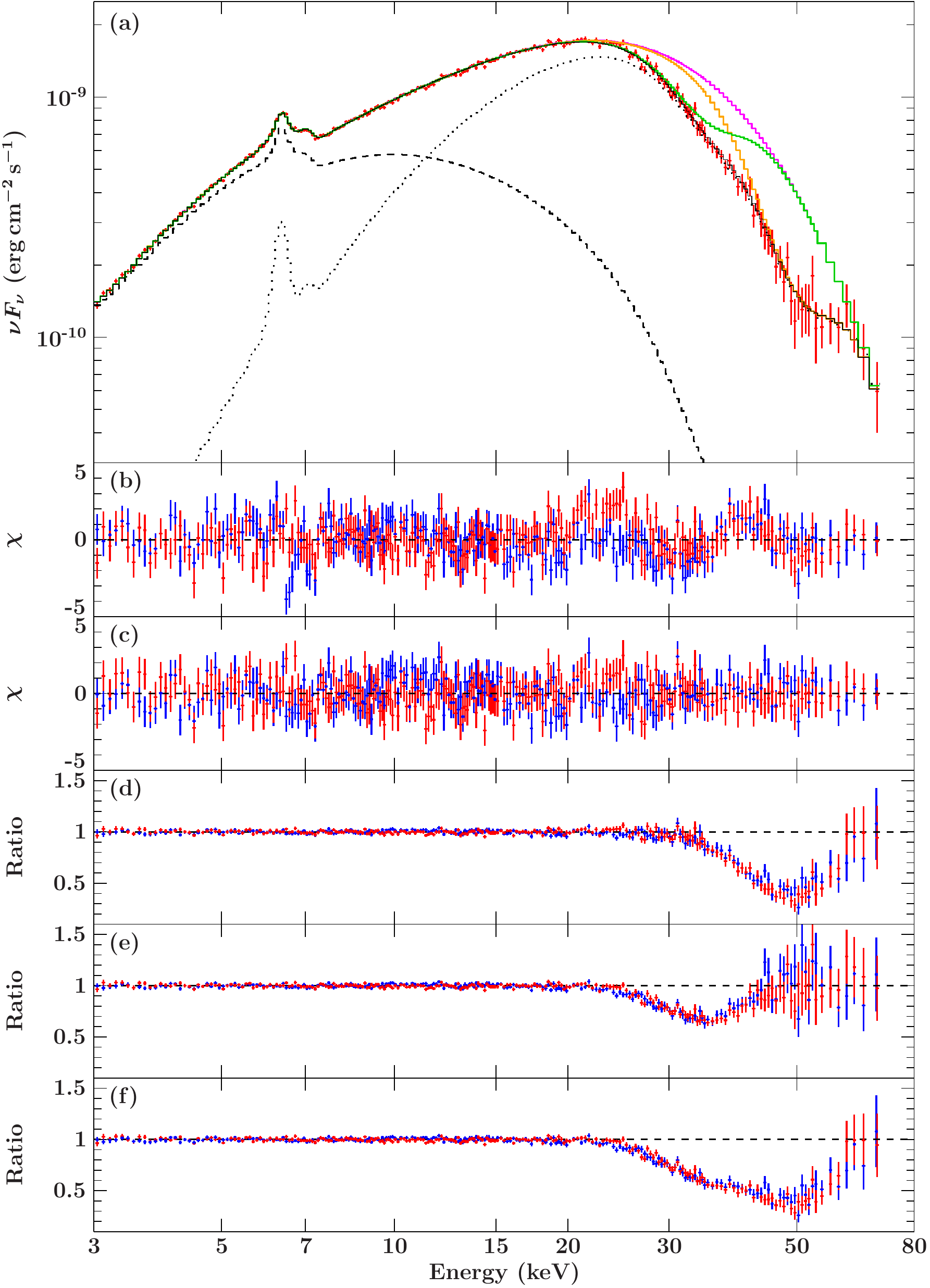}
\caption{Spectrum of \nustar observation 2. (a) Data from \nustar/FPMA (FPMB not shown for clarity), together with the best-fit \texttt{NPEX} model (black). We also show the two components  of the \texttt{NPEX} model separately, indicating the soft component with a dashed line and the harder component (for which the photon-index is fixed to $-2$) with a dotted line. Additionally, we show the model with each CRSF in turn removed (orange and green) and without both of them (magenta). (b) Residuals in terms of $\chi$ for the best-fit model with only 1 CRSF. (c) Residuals to the best-fit model with 2 CRSFs (d) Residuals in terms of data-to-model-ratio without the 50\,keV CRSF. (e) Residuals without the 35\,keV CRSF. (f) Residuals without both CRSFs. In this panel the non-symmetric shape of the residuals can be clearly seen. }
\label{fig:specobs2}
\end{center}
\end{figure}

None of the continuum models provide an acceptable fit, see Table~\ref{tab:obs2all_gabs1} in the appendix, where we also give the best-fit parameters of all tested models (including a gain-shift component as described below).
We found that the best fit is obtained by using the  \texttt{NPEX} model, resulting in $\redchi=1.27$ for 457 degrees of freedom (dof). 
Still, as shown in Fig.~\ref{fig:specobs2}(b) even this model leaves very wavy residuals between 20--50\,keV and requires a very broad CRSF ($\sigma > 10$\,keV). The wavy residuals may indicate that the CRSF shape is not captured correctly by the one \texttt{gabs} component \citep[\textit{cf.} Cep~X-4,][]{cepx4} or that a second line is present. 

We therefore added a second multiplicative Gaussian line, CRSF$_2$, also described by a \texttt{gabs} model. This approach resulted in a significantly improved fit with $\redchi=1.05$ for 454 dof (i.e., $\Delta\chi^2 = 103$ for 3 additional parameters). We find a strong line around 50\,keV, as well as a much shallower feature around 35\,keV.

However, in this fit small discrepancies between the FPMA and FPMB spectra become visible, beyond the small normalization difference of $\sim$3\%. This discrepancy appears mainly as a small energy off-set between the two modules, in particular around the \feka line. We therefore add an ad-hoc gain-shift (GS) to the FPMA energies, for which we find a best-fit intercept value around $\sim$20\,eV, i.e., half a bin of the native \nustar binning. Note that this is a constant, energy-independent shift applied to all model channels within ISIS. This small shift is only necessary due to the extremely high statistics in our data and is about a factor of 20 smaller than the full-width, half-maximum energy resolution of the detectors \citep{harrison13a}. It is also below the typical absolute uncertainties of the \nustar gain, which are on the order of 40\,eV in the iron line region \citep{nustarcalib}. Note that the applied gain-shift provides limits on how well the line energies can be reconstructed on an absolute scale, even though the statistical uncertainties on, e.g., the \feka energy are much smaller.

While the \npex continuum provided the statistically best-fit overall, using two independent \texttt{gabs} components to model the complex region between 30--60\,keV resulted in a statistically good fit for \fdcut and \texttt{HIGHECUT} model as well, but not  for the \texttt{cutoffpl} model (see Table~\ref{tab:obs2all_gabs2}). This is in agreement with previous hard X-ray studies of \gx, which also found a good description of the data only with more complex continuum models \citep[e.g.,][]{labarbera05a, suchy12a}. For all models with statistically acceptable fits the parameters of the two CRSF lines agree: the stronger one is located at $\sim$50\,keV, while the weaker one is located at $\sim$35\,keV. 
We note, however, that in the \fdcut model, we had to fix the strength of the high energy CRSF at the upper limit or 30, as it was unconstrained in the fit. We list the best-fit parameters for all models in Tab.~\ref{tab:phasavg}.

As can be seen in Fig.~\ref{fig:specobs2} the CRSFs have significant overlap, due to their relatively large widths. They remove over half of the continuum flux between 25--70\,keV, raising the question how realistic the estimated continuum is.  We therefore tried to fit also other continuum models without CRSFs, including a broken power-law model as well as a \texttt{comptt} model. However, we do not find another statistically acceptable description of the very sharp turnover around 25\,keV, together with the relatively straight continuum above that energy, modulated by wavy structure and the hardening above 60\,keV. 

The relative width $\sigma/E$ of the CRSF is in line with what is expected from theoretical models and thermal line broadening \citep{schwarm17b}, with values around 15\% for both lines. The similar relative width of both lines is consistent with them being formed in the same plasma and gives us an indication about their physical origin.

We also checked if the CRSF is better described by a single line with a Lorentzian optical depth profile (using the CYCLABS model in XSPEC), however, we do not find an acceptable fit with only one line. This finding can be understood due to the fact that the lines are broad, and the difference between a Gaussian and Lorentzian optical depth profile is negligible in this regime. 

While the spectrum of \gx is rather complex at soft energies, due to the strong absorption and large contribution from fluorescence lines, no such features, besides the CRSF, are known above 10\,keV. That is, the shape of the hard X-ray continuum, caused by Comptonization of seed photons in the hot electron gas of the accretion column, is expected to be very smooth. Any deviation from such a smooth continuum should therefore only be caused by effects from the magnetic field, i.e., CRSFs.

To test the assumption that the continuum above 10\,keV is smooth, we applied the \texttt{compmag} model \citep{farinelli12a}. This model implements a physical model of the accretion column  based on theoretical calculations by \citet{becker07a}. 
With this physical continuum model alone, we do not find a good fit.  However, when adding a thermal component in the form of a multi-temperature black-body (\texttt{diskbb}), we can find a very good description of the data, if, and only if, we also include two CRSFs ($\redchi = 1.02$ for 451 dof). The parameters of the CRSFs are consistent with the ones we find using the \texttt{NPEX} model (see Tab.~\ref{tab:obs2all_gabs2}). Using only one CRSF does not lead to an acceptable fit (Table~\ref{tab:obs2all_gabs1}). The parameters we find for the \texttt{compmag} model produce a continuum that is very smoothly rolling over at high energies, which is very closely mimicked by the \npex model.

To test the significance of the detection of a second CRSFs we run Monte Carlo simulations. We simulate 10\,000 spectra, based on our best-fit with only one CRSF using the \npex continuum (for the sake of computing speed, we do not use the physical \texttt{compmag} model, however, as  \npex produced an almost identical continuum shape, the results will also hold up for the physical modeling). For each simulated spectrum, we draw the model parameters from a Gaussian distribution around the best-fit values. We choose the width of the distribution to match the 90\% uncertainties of the spectral parameters and apply Poisson statistics before folding it through the \nustar responses. We then apply the same binning schemes as for the real data and fit the simulated spectra with an NPEX continuum model, once with one CRSF and then again with two CRSFs. We  compare the $\chi^2$ between these two models, which  allows us to sample how likely it is to find an improvement in terms of $\chi^2$ as large as in the real data, which translates to a false positive rate of an improvement of that strength.
We find that only for 1 of 10.000 simulated spectra the models show  an improvement as large or larger than the real data. This means, the false positive rate is $\leq$0.01\% and that the feature in the real data is statistically significant at $\geq99.9\%$. 

Another test of significance is to apply the  Akaike Information Criterion \citep[AIC,][]{akaike74a}, which takes the degrees of freedom in each model into account. In particular we use  $\Delta$AIC, the difference in AIC between the fit with one CRSF to the one with two CRFS. From that we find a false positive rate of only $10^{-19}$. While this is much lower than inferred from the simulations,  the  AIC probability at these extreme ends of the distribution should be treated with some caution. Furthermore, we  expect that the false positive rate in our simulations to go down further if running more simulated spectra, however, the required numbers is computationally prohibitive.

\citet{labarbera05a} used \sax data to perform a detailed study of the \gx, and did not find evidence for a secondary line in \sax data in apparent contradiction to our results. In fact, they do not detect any significant line during their observation at a similar orbital phase (0.62--0.65) as the \nustar data presented here. However, the authors state that this non-detection is likely due to a lack of signal in the data.

To check if \sax could have picked up on the second line during a theoretical observation in a similar state as the \nustar observation with high enough \snr, we simulated 10\,000 \sax spectra of the MECS, HPGSPC, and PDS instruments, based on responses and background data as published in the CALDB from May 1999. We use exposure times of 100\,ks in MECS and 50\,ks in HPGSPC and PDS, the latter being lower due to the rocking of the instruments. We use our best-fit \texttt{NPEX} model with two CRSFs as input, again drawing randomly from a Gaussian distribution around the best-fit values. We then compare the improvement between a model with two lines to a model with one line, based on the AIC. We find that in about 50\% of all simulated spectra the fit is significantly improved ($\Delta$AIC > 5) when using two lines. This result shows that the second line is at the edge of the \sax sensitivity, and is consistent with the non-detection in the existing data, which show a lower \snr than our simulated spectra (which also implicitly  assume a perfect knowledge of the responses). In comparison to \inte and \suz, \sax had the best detectors to measure the line profile, given its continuous coverage and good energy resolution. We can therefore confidently say, that the non-detection of a secondary line in both \suz and \inte can be ascribed to the lack of data quality in these instruments (and see Sect.~\ref{susec:suz} for a comparison to archival \suz data).

The \texttt{NPEX} model essentially consists of two \texttt{cutoffpl}s with the same folding energy, but with one component having a positive and the other a negative photon-index. We froze the index of the negative component to $\Gamma$=$-2$ (note that in the definition of \texttt{cutoffpl} $\mathcal{F} \propto E^{-\Gamma}$). Therefore, one of these components  dominates the hard energies, while the other component dominates the soft energies. We calculate the flux for each of these components separately, which will give us another indicator for the spectral hardness and also plot the two components in Fig.~\ref{fig:specobs2}\textit{a}. We will refer to this fluxes as ``hard'' and ``soft'', respectively.

\begin{sidewaystable*}
\caption{Best-fit model parameters for the phase-averaged fits.}\label{tab:phasavg}
\centering
\begin{tabular}{r|lll|lll|l}
\hline\hline
& \multicolumn{3}{|c}{Observation 1} & \multicolumn{3}{|c}{Observation 2} & Suzaku \\
 Parameter & NPEX 1 & FDcut 1 & HighECut & NPEX & FDCut & HighECut & NPEX \\\hline
 $ N_\text{H,1}~(10^{22}\,\text{cm}^{-2})$ & $39.8\pm1.5$ & $39.6\pm1.3$ & $39.9^{+1.3}_{-1.4}$ & $29.8\pm1.6$ & $25.7\pm1.6$ & $26.3\pm1.6$ & $50.9\pm1.3$ \\
 $ f$ & $0.910\pm0.006$ & $0.9201^{+0.0010}_{-0.0041}$ & $0.921^{+0.005}_{-0.004}$ & $0.828\pm0.009$ & $0.829^{+0.012}_{-0.011}$ & $0.830^{+0.012}_{-0.010}$ & $0.9926^{+0.0009}_{-0.0010}$ \\
 $ \mathcal{F}~(10^{-9}\,\text{erg}\,\text{cm}^{-2}\,\text{s}^{-1})$\tablefootmark{a}$$ & $1.757\pm0.012$ & $1.760\pm0.011$ & $1.762\pm0.011$ & $2.583\pm0.016$ & $2.545\pm0.014$ & $2.550^{+0.015}_{-0.014}$ & $6.58\pm0.08$ \\
 $ A_\text{2}$\tablefootmark{b}$$ & $\left(6.8^{+0.9}_{-0.8}\right)\times10^{-5}$ & --- & --- & $\left(1.47\pm0.12\right)\times10^{-4}$ & --- & --- & $\left(5.5\pm0.6\right)\times10^{-4}$ \\
 $ \Gamma$ & $0.788\pm0.030$ & $1.454^{+0.021}_{-0.025}$ & $1.467^{+0.020}_{-0.022}$ & $0.840\pm0.029$ & $1.308^{+0.020}_{-0.021}$ & $1.325\pm0.020$ & $0.67^{+0.13}_{-0.12}$ \\
 $ E_\text{cut}~(\text{keV})$ & --- & $45.5^{+4.1}_{-2.2}$ & $37.3^{+0.8}_{-1.0}$ & --- & $44.2\pm2.0$ & $21.87^{+0.38}_{-0.29}$ & --- \\
 $ E_\text{fold}~(\text{keV})$ & $6.52^{+0.35}_{-0.23}$ & $6.3^{+1.0}_{-1.1}$ & $8.89^{+0.85}_{-0.20}$ & $6.10^{+0.17}_{-0.13}$ & $5.3^{+0.6}_{-0.7}$ & $11.8^{+1.0}_{-0.6}$ & $5.82^{+0.16}_{-0.14}$ \\
 $ \text{EW}_{\text{K}\alpha}~(\text{eV})$ & $\left(1.08^{+0.08}_{-0.07}\right)\times10^{2}$ & $\left(1.49^{+0.09}_{-0.08}\right)\times10^{2}$ & $\left(1.10\pm0.07\right)\times10^{2}$ & $\left(1.34\pm0.06\right)\times10^{2}$ & $\left(1.49^{+0.09}_{-0.08}\right)\times10^{2}$ & $\left(1.46^{+0.09}_{-0.08}\right)\times10^{2}$ & $\left(1.38\pm0.05\right)\times10^{2}$ \\
 $ E_{\text{K}\alpha}~(\text{keV})$ & $6.352^{+0.005}_{-0.032}$ & $6.355795^{+0.000011}_{-0.034438}$ & $6.3548^{+0.0011}_{-0.0335}$ & $6.337^{+0.019}_{-0.008}$ & $6.333\pm0.009$ & $6.333\pm0.009$ & $6.410^{+0.009}_{-0.006}$ \\
 $ \sigma_{\text{K}\alpha}~(\text{keV})$ & $\left(1.6^{+49.9}_{-1.6}\right)\times10^{-3}$ & $\le0.07$ & $\le0.06$ & $0.022^{+0.033}_{-0.022}$ & $0.066^{+0.022}_{-0.028}$ & $0.060^{+0.023}_{-0.031}$ & --- \\
 $ E_{\text{K}\beta}~(\text{keV})$ & --- & --- & --- & $7.05^{+0.20}_{-0.10}$ & $7.04^{+0.05}_{-0.16}$ & $7.04^{+0.06}_{-0.16}$ & $7.087^{+0.015}_{-0.025}$ \\
 $ \text{EW}_{\text{K}\beta}~(\text{eV})$ & $-6\pm6$ & $10\pm5$ & $-4^{+6}_{-5}$ & $10\pm5$ & $10\pm5$ & $10\pm5$ & $23\pm4$ \\
 $ E_\text{CRSF,1}~(\text{keV})$ & $50.6^{+2.1}_{-1.7}$ & $50.2714^{+0.0007}_{-1.7640}$ & $48.9^{+3.4}_{-2.2}$ & $49.6^{+1.3}_{-1.2}$ & $49.2\pm1.2$ & $50.4^{+1.8}_{-1.4}$ & --- \\
 $ \sigma_\text{CRSF,1}~(\text{keV})$ & $8.8^{+1.2}_{-2.3}$ & $8.0090^{+0.1674}_{-0.0011}$ & $9.0^{+1.4}_{-2.3}$ & $7.8^{+2.3}_{-1.5}$ & $7.0^{+0.7}_{-0.6}$ & $7.4^{+2.1}_{-1.9}$ & --- \\
 $ d_\text{CRSF,1}~(\text{keV})$ & $20^{+9}_{-7}$ & --- & $27.6^{+2.5}_{-8.9}$ & $20^{+9}_{-5}$ & --- & $13^{+8}_{-5}$ & --- \\
 $ E_\text{CRSF,2}~(\text{keV})$ & $34.7^{+2.1}_{-1.4}$ & $35.7^{+2.2}_{-1.7}$ & $31.3^{+6.8}_{-1.3}$ & $34.5^{+1.6}_{-1.4}$ & $35.6\pm1.3$ & $35.1^{+1.6}_{-1.2}$ & $35.1^{+0.8}_{-0.7}$ \\
 $ \sigma_\text{CRSF,2}~(\text{keV})$ & $5.0^{+1.1}_{-1.2}$ & $6.3199^{+0.8459}_{-0.0023}$ & $5.0^{+2.1}_{-0.9}$ & $5.1\pm0.8$ & $6.5\pm0.5$ & $3.5^{+1.0}_{-0.9}$ & $6.7^{+0.9}_{-0.8}$ \\
 $ d_\text{CRSF,2}~(\text{keV})$ & $3.8^{+3.3}_{-2.1}$ & $11.2906^{+7.5606}_{-0.0011}$ & $4.6^{+20.4}_{-2.0}$ & $5.1^{+2.6}_{-2.5}$ & $16\pm5$ & $1.5^{+1.1}_{-0.7}$ & $8.9^{+2.8}_{-2.0}$ \\
 $ CC_\text{B}$ & $1.040\pm0.004$ & $1.040\pm0.004$ & $1.040\pm0.004$ & $1.0373\pm0.0030$ & $1.0372\pm0.0030$ & $1.0373\pm0.0030$ & $0.97^{+0.05}_{-0.04}$ \\
 $ \text{GS~(eV)}$ & $0.036^{+0.009}_{-0.010}$ & $0.035\pm0.009$ & $0.036\pm0.009$ & $0.024\pm0.009$ & $0.019\pm0.009$ & $0.020\pm0.009$ & --- \\
$\mathcal{L}~(10^{36}$\,erg\,s$^{-1}$)\tablefootmark{c} &  $1.892\pm0.013$ & $1.895\pm0.012$ & $1.898\pm0.012$ & $2.782\pm0.017$ & $2.740^{+0.016}_{-0.015}$ & $2.745^{+0.016}_{-0.015}$ & $7.08\pm0.08$ \\
\hline$\chi^2/\text{d.o.f.}$   & 552.65/454& 556.28/455& 549.93/452& 456.91/454& 559.94/454& 533.75/451& 334.46/277\\$\chi^2_\text{red}$   & 1.217& 1.223& 1.217& 1.006& 1.233& 1.183& 1.207\\\hline
\end{tabular}
\tablefoot{\tablefoottext{a}{unabsorbed  flux between 5--50\,keV}
\tablefoottext{b}{in ph\,keV$^{-1}$\,cm$^{-2}$\,s$^{-1}$~at 1\,keV}
\tablefoottext{c}{luminosity between 5--50\,keV for a distance of 3.0\,kpc}
}
\end{sidewaystable*}

\subsection{\nustar observation 1}

Having established a good description of the spectrum of observation~2 in the previous section, we use the same models for the fainter data of observation 1. We use the same binning scheme as for observation 2 and find again an acceptable fit only when using two independent CRSFs.  The statistical quality of the fit for the \npex model is worse than in observation~2 and no longer clearly better than the \fdcut or \highe model (Table~\ref{tab:obs1all_gabs2}). All three models resulted in statistically acceptable fits, and we find that the CRSF parameters are independent of the chosen continuum model (Tab.~\ref{tab:phasavg}). The gain-shift intercept is larger in observation 1 than in observation 2, with a best-fit value around 36\,eV, which is still below the uncertainty of the energy calibration of \nustar. 

We also tried to model this observation with one CRSF only, but find that the improvement when adding a second line is statistically significant for all continua (Table~\ref{tab:obs1all_gabs1}). The smallest improvement is found for the \highe model, which still gives us a $\Delta$AIC value of 12.3, which corresponds to a significance of $>99\%$.

We find that the parameters of the CRSFs are consistent with the parameters found in observation~2, in particular with respect to the energy. We therefore do not find evidence for a luminosity dependence of the CRSF energy, however, as the flux is only about 30\% lower compared to observation~2 any correlation is likely hidden within the uncertainties.

\begin{figure}
\begin{center}
\includegraphics[width=0.95\columnwidth]{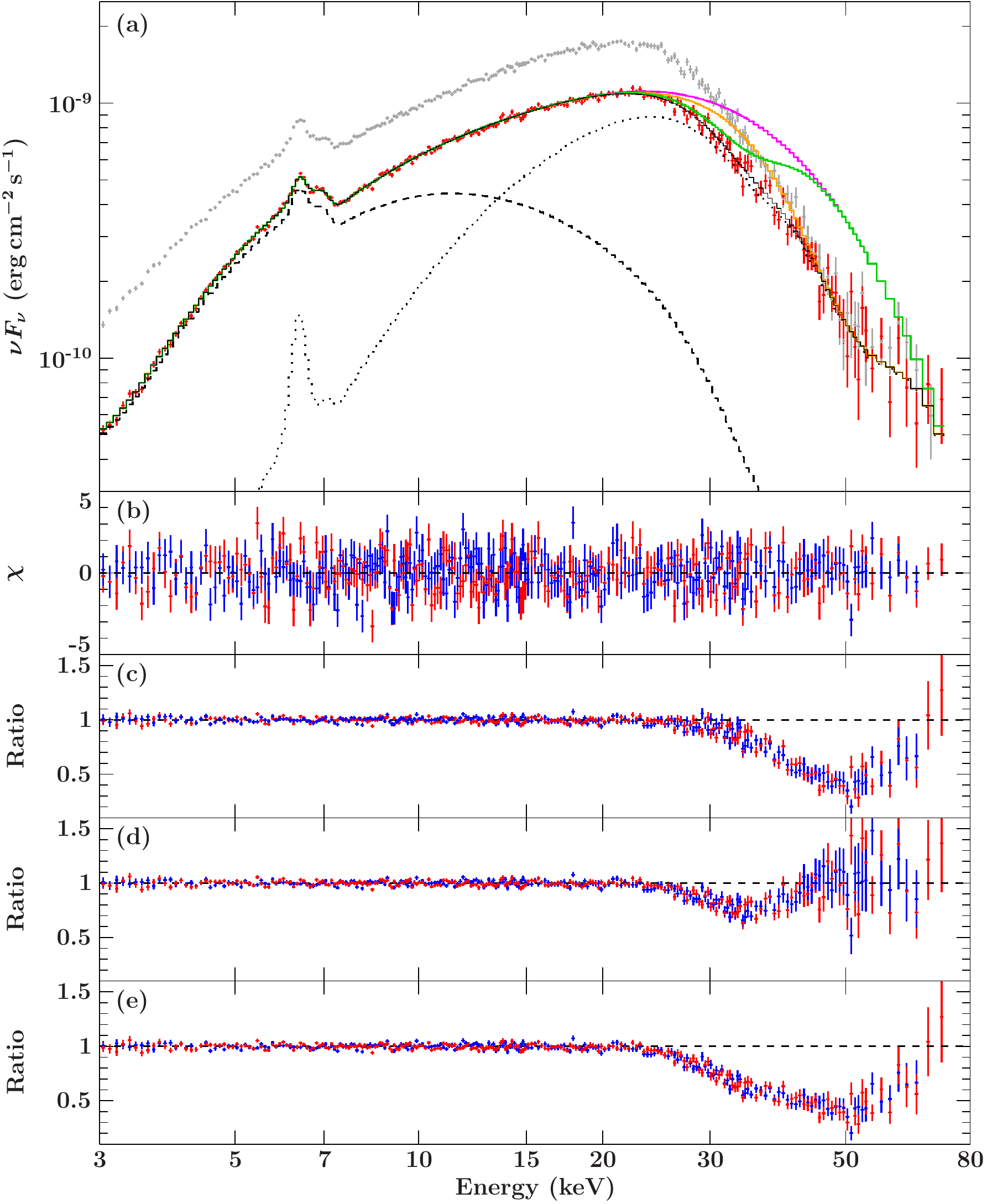}
\caption{Same as Fig.~\ref{fig:specobs2}, but for observation~1. (a) Additionally to the \nustar/FPMA data and best-fit NPEX model, we show the spectrum of observation~2 in gray for comparison. (b) Residuals in terms of $\chi$ for the best-fit model with 2 CRSFs. (c) Residuals in terms of data-to-model-ratio without the 50\,keV CRSF. (d) Residuals without the 35\,keV CRSF. (e) Residuals without both CRSFs. 
}
\label{fig:specobs1}
\end{center}
\end{figure}

In Fig.~\ref{fig:specobs1} we show the spectra of both observations. As can be seen the increased flux of observation~2 mainly occurs below $\sim$40\,keV, while above the spectra are very similar. This is reflected in the similar spectral parameters we find in both observations, where the only significant difference is the absorption column and the covering fraction.

\subsection{Comparison to \suz}
\label{susec:suz}

\suz observed \gx twice, as described by \citet{suchy12a}. Here we re-analyze the longer of the two observations, ObsID 4030444020, taken on 2009, January 05 (Tab.~\ref{tab:obslog}). \citet{suchy12a} found a very good description of the spectrum with just one CRSF at around 35\,keV and our goal is to see if the best-fit model we find for \nustar, which requires two CRSFs, is also able to describe the \suz data. As \citet{suchy12a} discuss, the spectral parameters are somewhat dependent on the combination of XIS detectors  used, due to calibration difference between the front-illuminated chips (XIS0 and XIS3) and back-illuminated chip (XIS1). As our main focus is on the hard X-ray spectrum, we will only use data from XIS1 and PIN here for simplicity.

We re-binned the XIS data within ISIS following the scheme described by \citet{nowak11a}, which follows the energy resolution of the detector, and requiring at least a \snr of 8 per bin. We restricted the energy range of XIS to be between 3--10\,keV to reduce the influence of the very complicated soft X-ray spectrum on the statistical quality of the fit. We re-binned the PIN data to \snr of 5 per bin and used them between 15--80\,keV. Fluxes are given relative to PIN and we added a cross-calibration constant for the XIS data, to account for differences in the absolute flux calibration between these detectors.

To model the data, we start with the best-fit NPEX model for \nustar observation 2. As the \snr of PIN above 50\,keV is very low, we chose to fix the high energy CRSF to the parameters found in \nustar. Due to the much higher spectral resolution of XIS compared to \nustar, we additionally added a \nika line at around 7.5\,keV and a Ca~K$\alpha$ line at 3.7\,keV.

\begin{figure}
\begin{center}
\includegraphics[width=0.95\columnwidth]{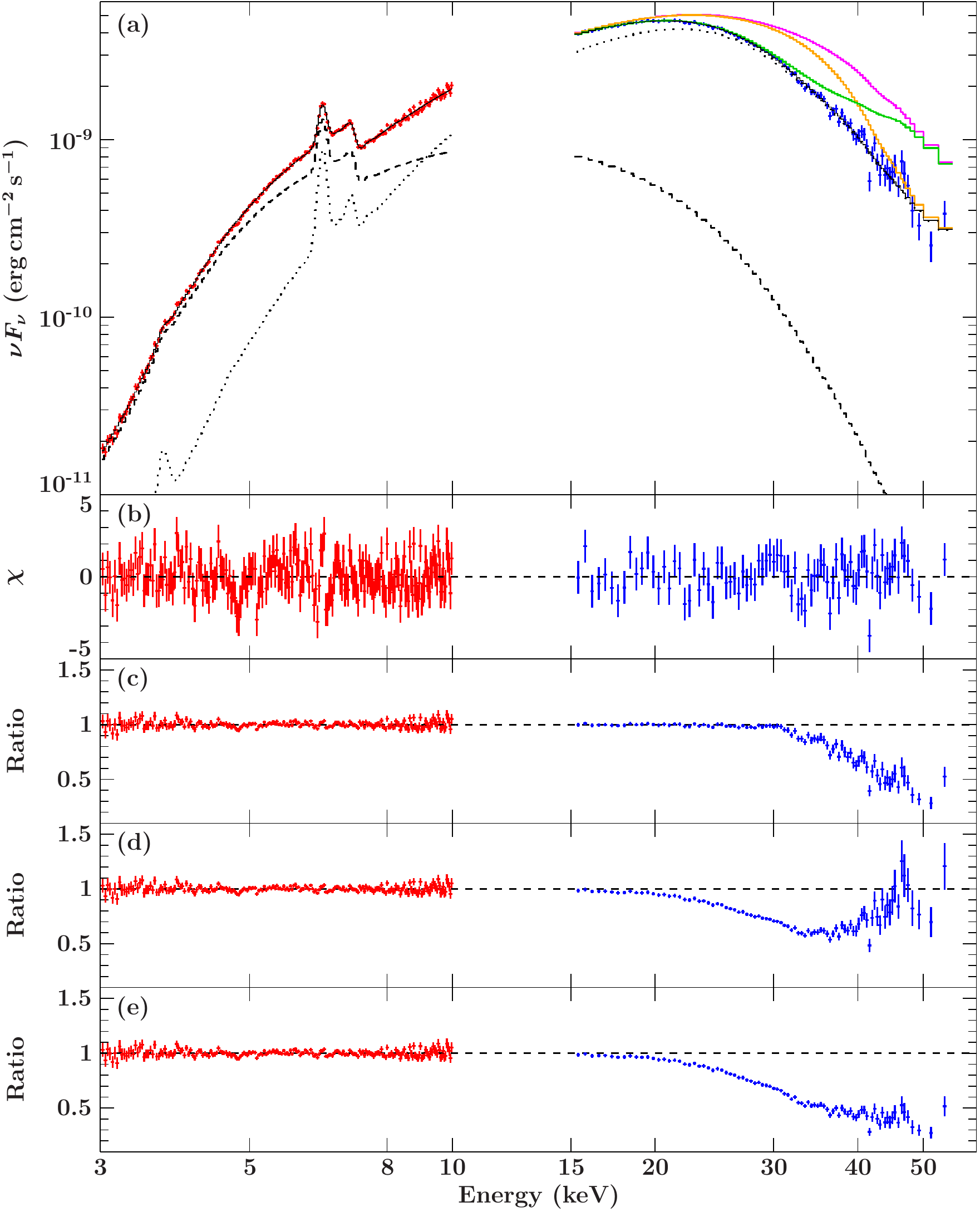}
\caption{Same as Fig.~\ref{fig:specobs2} but for the \suz spectrum. (a) XIS\,1 data are shown in red, PIN data are shown in blue. (b) Residuals in terms of $\chi$ for the best-fit model with 2 CRSFs. (c) Residuals in terms of data-to-model-ratio without the 50\,keV CRSF. (d) Residuals without the 35\,keV CRSF. (e) Residuals without both CRSFs. }
\label{fig:specsuz}
\end{center}
\end{figure}

We find that this model describes the data very well, with similar parameters for the low energy CRSF as we found in the \nustar data (Tab.~\ref{tab:phasavg}). However, the strength of the CRSF is somewhat stronger, and the feature is removing significant flux from the implied continuum (Fig.~\ref{fig:specsuz}). From the \suz data alone our model therefore might appear somewhat unphysical, as the CRSFs dominate the spectral shape above $\sim$20\,keV. However, as we have seen that this model is clearly required by the \nustar data, the fact that we also find an acceptable description to the much brighter \suz data is a further indication that we use the correct description. 
We note that the line at 35\,keV found by \citet{suchy12a} has very similar parameters to the line we find here, indicating in their model the 50\,keV feature was adequately modeled by the different continuum used.
.

\section{Phase-resolved spectroscopy}
\label{sec:phasres}

\subsection{Pulse profiles}
\label{susec:pp}

For the timing and pulse phase-resolved analysis, we shifted all time information to the solar barycenter using the task \texttt{barycorr}. We also corrected for the binary motion, using the ephemeris by \citet{koh97a}. To find the pulse period, we folded the events and light-curve on test periods in the vicinity of the known 685\,s period and searched for the folded profile with the largest deviation from a constant \citep[i.e., using the epoch folding technique,][]{leahy83a}. To determine the uncertainty on the period, we simulated 1000 light-curves based on the pulse profile and performed the same search. The uncertainty is the $1\sigma$ interval of the distribution of the resulting periods. To include the strong pulse-to-pulse variations of \gx, we added a noise term to these simulated light curves to obtain the same variance as in the original light curve \citep{davies90a}. We find $P_\text{Obs. 1} = 686.61\pm0.03$\,s and $P_\text{Obs 2} = 685.83 \pm 0.05$\,s, indicating significant torque transfer onto the neutron star between these two epochs. The measured periods are in good agreement to those measured at the respective times by \fermi/GBM\footnote{\url{https://gammaray.nsstc.nasa.gov/gbm/science/pulsars/lightcurves/gx301m2.html}} \citep{finger09a}.

\begin{figure}
\begin{center}
\includegraphics[width=0.95\columnwidth]{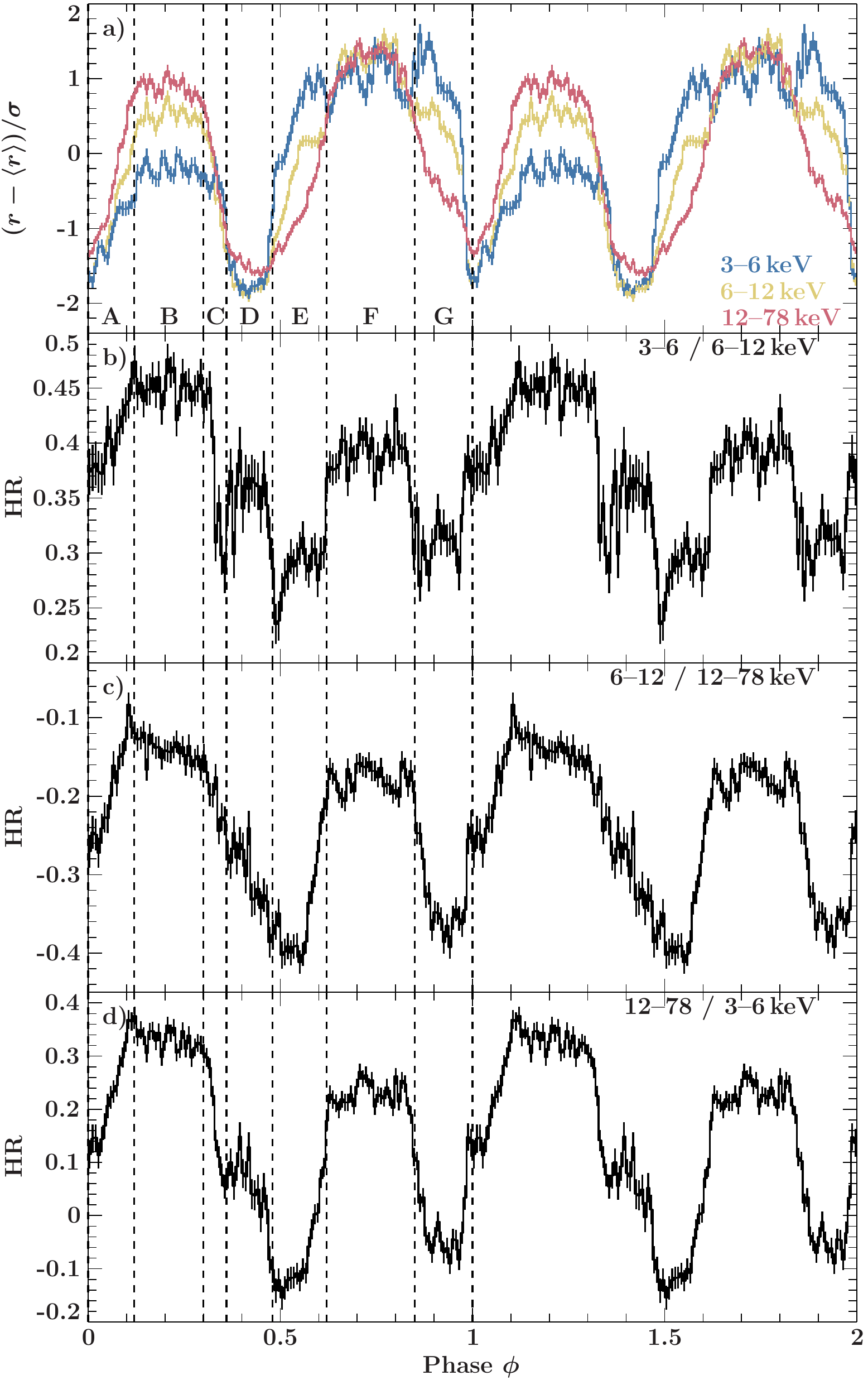}
\caption{(a) Pulse profiles of observation 2 in three different energy bands: 3--6\,keV (blue), 6--12\,keV (yellow), and 12--78\,keV (red).  The profiles were normalized by subtracting the respective mean rate and dividing by the respective standard deviation to allow for easier comparison of the shape between energy bands. The lower panels show the hardness ratio between each pair of these pulse profiles. The hardness ratio is calculated as $(H-S)/(H+S)$, where $H$ is the profile in the harder band, and $S$ the profile in the softer band. (b) Hardness ratio between the 3--6\,keV and 6--12\,keV profile. (c) Hardness ratio between the 6--12\,keV and 12--78\,keV profile. (d) Hardness ratio between the 3--6\,keV and 12--78\,keV profile. The dashed lines indicate the phases selected for phase-resolved spectroscopy. }
\label{fig:pp}
\end{center}
\end{figure}

The pulse profiles of observation 2 in three energy bands are shown in Fig.~\ref{fig:pp}, together with the pairwise hardness ratios of these profiles. A very strong energy dependence is evident, i.e., the main peak (around phase $\phi = 0.75$) becomes significantly narrower at higher energies, while the relative strength of the secondary peak (around $\phi = 0.2$) increases. This evolution agrees well with the pulse profiles observed by \xte, \sax, and \suz \citep{kreykenbohm04a, labarbera05a, suchy12a}.
The energy dependence is strongest at low energies, while above $\sim$12\,keV it is much less pronounced. We therefore chose to combine the high-energy data into one profile to increase the \snr. 

\begin{figure}
\begin{center}
\includegraphics[width=0.95\columnwidth]{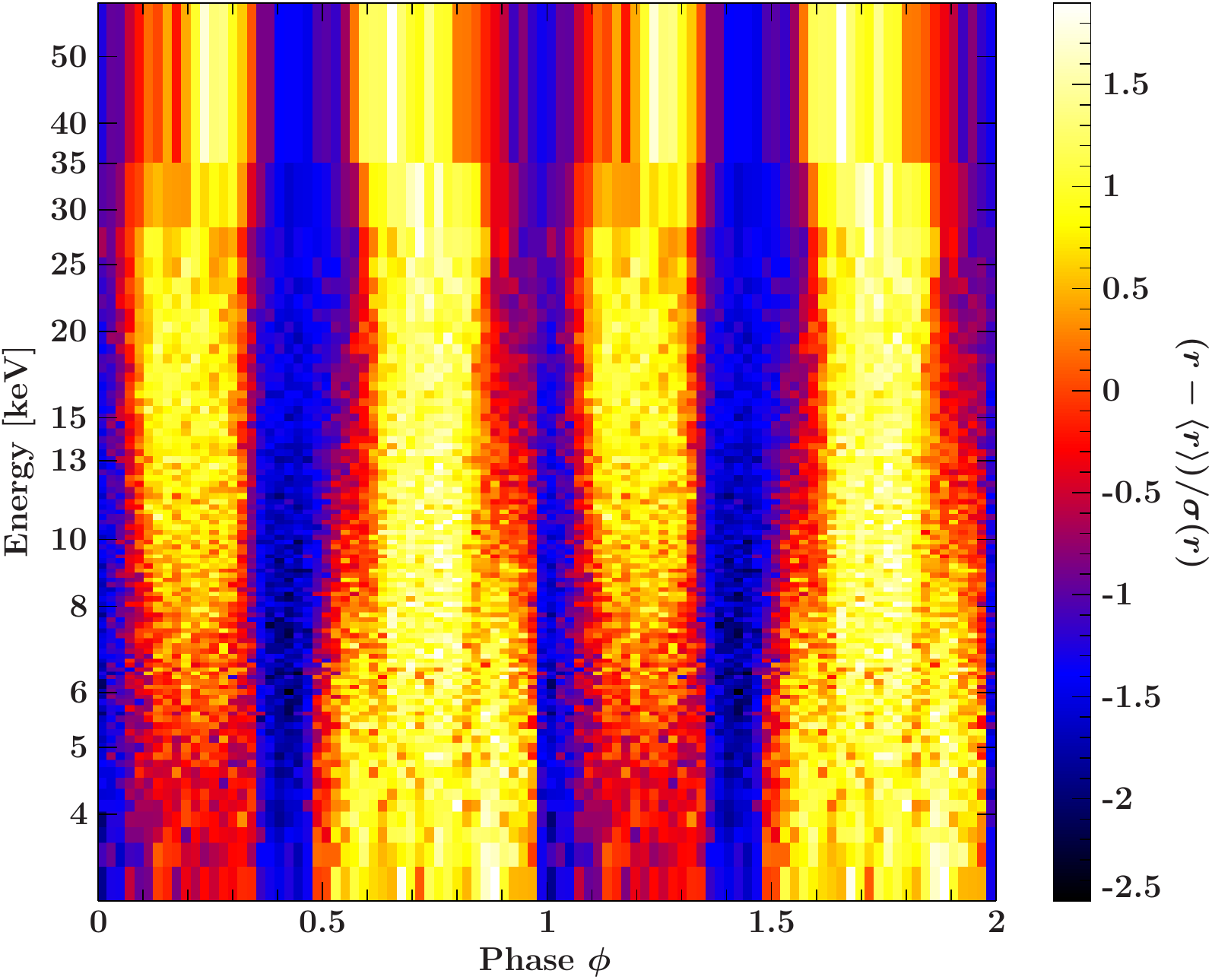}
\caption{Color-coded energy resolved pulse profile map. The $x$-axis indicates phase (repeated once for clarity) and the $y$-axis energy in keV. The color indicates the relative strength in each energy band according to the scale on the right, normalized by $\left(r-\left<r\right>\right)/\sigma(r)$, where $r$ is the rate in each bin, $\left<r\right>$ is the average count-rate over each energy band, and $\sigma(r)$ is the corresponding standard deviation.}
\label{fig:ppergmap}
\end{center}
\end{figure}

To highlight the energy dependence of the pulse profile we also show a color-coded map of the energy resolved profiles of observation 2 (Figure~\ref{fig:ppergmap}).
The energy bins were chosen so that each band  contains 7500 photons. From this binning it follows that the highest energy resolution is obtained around the \feka line at 6.4\,keV, where the highest count-rates are measured.
The different behavior as function of energy between the two main peaks can be clearly seen, with the secondary one (around phase $\phi=0.2$) being much weaker at low energies, but becoming similarly strong as the primary one at higher energies.

The energy resolved pulse profiles of observation 1 (not shown) look qualitatively the same, in particular the energy dependence is very similar. The gap between the primary and secondary peak is somewhat narrower in observation 1, but still clearly discernible.

To perform phase-resolved analysis, we split the data into seven phase-bins, indicated in Fig.~\ref{fig:pp}. These bins were selected to roughly cover intervals of constant hardness, and in particular to provide a very good \snr across the two peaks of the pulse profile where the spectrum seems to change only marginally as a function of phase. 
We label the phase-bins with letters A--G, where phase-bin B corresponds to the first peak and phase-bin F corresponds to the second peak.

\subsection{Phase-resolved analysis of observation 2}
\label{susec:phasres2}

To describe the spectra of the seven selected phase bins, we use the \texttt{NPEX} model, including two CRSF lines and the iron lines, as well as partially covering absorption. 
This is the same model as was found to describe the phase-averaged spectrum best (see Tab.~\ref{tab:phasavg}).
However, due to the reduced \snr in each spectrum, we had to fix the width $\sigma$ of the CRSF lines, as well as the width and energy of the \feka and \fekb lines to the best-fit values of the phase-averaged spectrum. After preliminary fits, we also decided to fix the folding energy $E_\text{fold}$ to the phase-average value, as we did not find a significant change of this parameter as function of phase and it has some degeneracy with the photon-index $\Gamma$. Therefore, we artificially describe all changes in the continuum shape by variations of $\Gamma$ and the ratio of the hard and soft continuum component.

\begin{figure}
\begin{center}
\includegraphics[width=0.95\columnwidth]{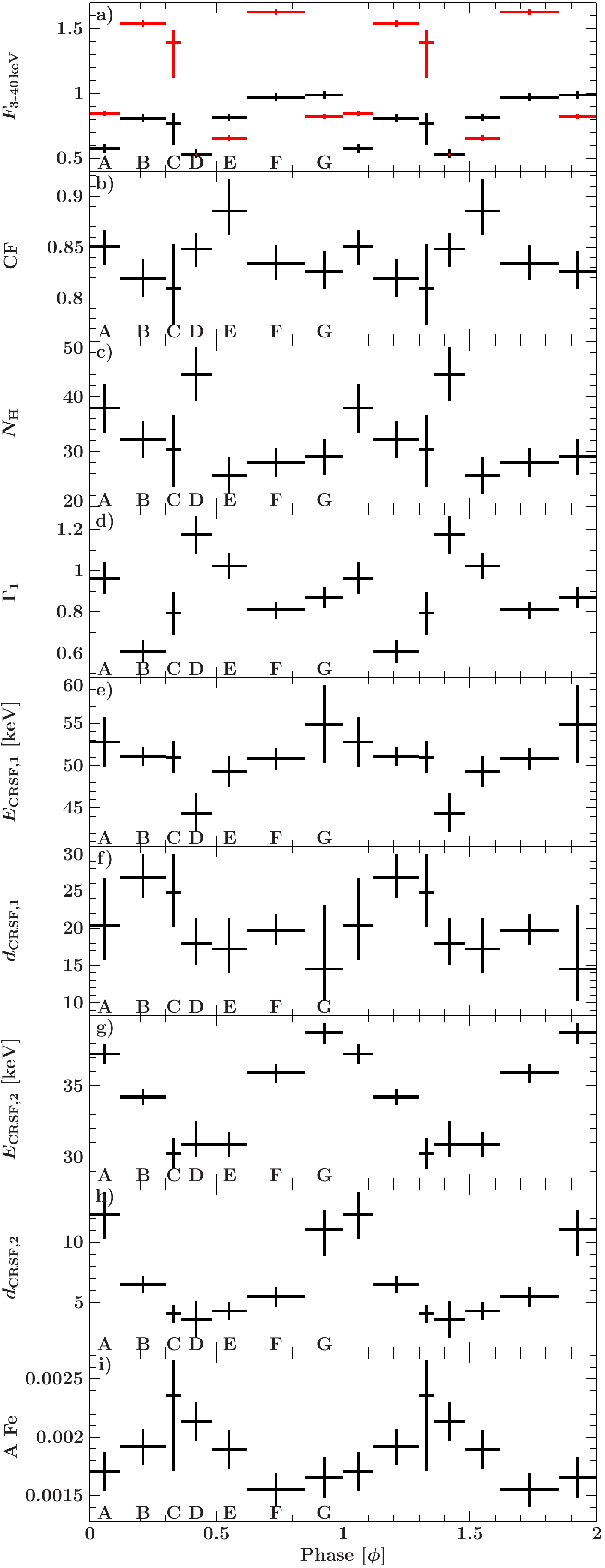}
\caption{(a) Flux between 3--40\,keV in keV\,s$^{-1}$\,cm$^{-2}$. Shown in black is the ``soft'' component, in red the ``hard'' component (see Sect.~\ref{sec:phasavg} for a description of the components). (b) Covering fraction. (c) Absorption column in $10^{22}$\,cm$^{-2}$. (d) Photon index. (e) Energy of CRSF$_1$. (f) Strength of CRSF$_1$. (g) Energy of CRSF$_2$. (h) Strength of CRSF$_2$. (i) Iron line flux in ph\,s$^{-1}$\,cm$^{-2}$.}
\label{fig:phasresobs2}
\end{center}
\end{figure}

The results of the phase-resolved fits are shown in Fig.~\ref{fig:phasresobs2} for observation 2.   As can be seen in the figure there is strong variability in all parameters. The flux of the hard and soft power-law follow very well the corresponding soft and hard pulse profiles (see Fig.~\ref{fig:pp}), in particular, the second peak in the hard flux (bin F) is much narrower than in the softer band. 

On top of this variation, the photon index also changes significantly, varying from around 0.6 in bin B to almost 1.2 in bin D (i.e., the deep minimum in the pulse profile). It is interesting to note that the hardness ratios do not take on extreme values during phase-bin D, indicating a complex interplay between the photon-index and the fluxes of the two components of the \npex model.

The CRSF around 35\,keV (CRSF$_2$) varies significantly in energy as well as in strength as function of pulse phase.  On the other hand all energies of CRSF$_1$ (around 50\,keV) are consistent with a constant besides the measurement in phase-bin D. As this bin is taken during the pulse minimum, we do not have data above 50\,keV, therefore the lower CRSF energy might be a systematic effect. To check that, we froze the energy of the CRSF in this phase-bin to the value from the phase-averaged fit, i.e., 49.6\,keV. With this, we find a statistically slightly worse fit $\Delta\chi^2=11$, while all other parameters did not change significantly. 

The variation of the parameters of the low-energy CRSF (energy and strength)  are only single peaked, in contrast to the double peaked pulse profile. In particular, the extrema do not line up with any of the peaks in the pulse profile, but seem to be shifted to later phases. 

\begin{figure}
\begin{center}
\includegraphics[width=0.95\columnwidth]{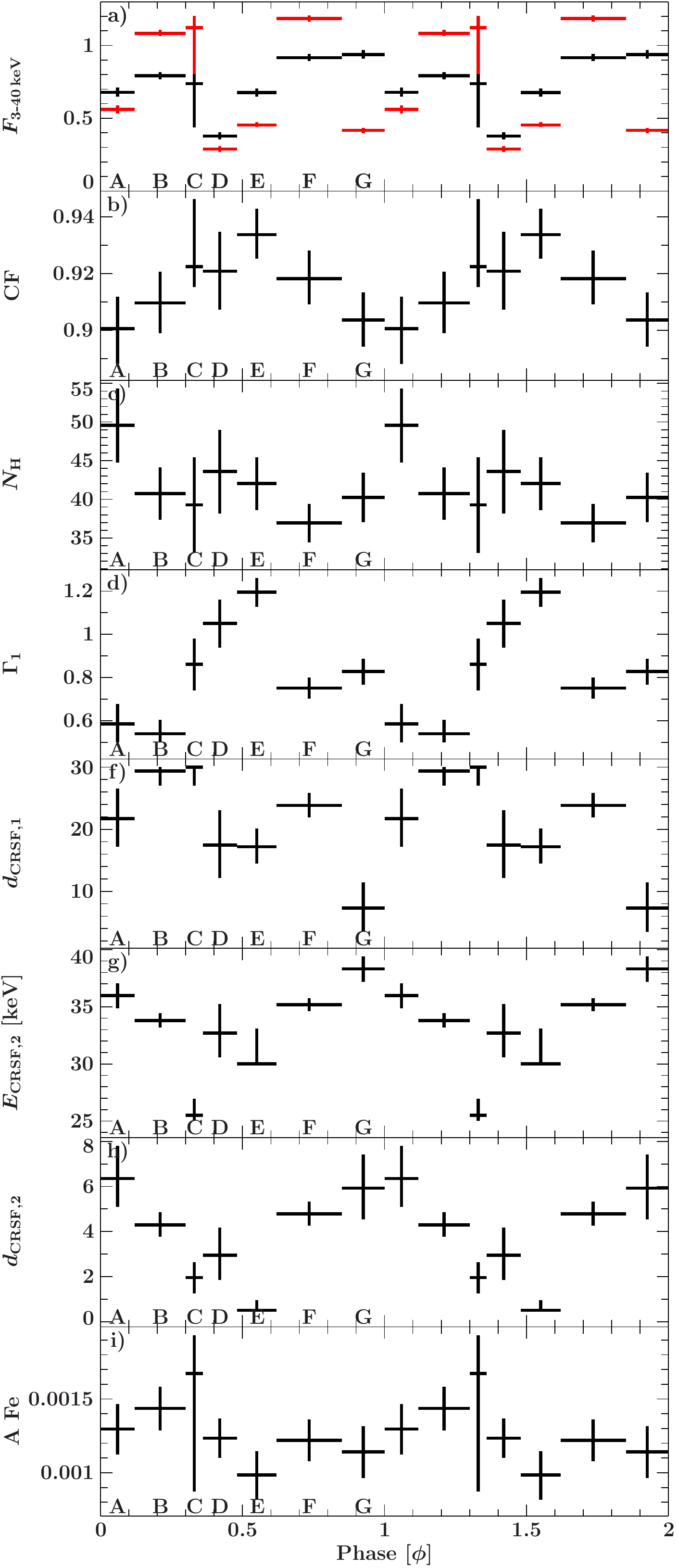}
\caption{Same as Fig.~\ref{fig:phasresobs2}, but for observation 1.}
\label{fig:phasresobs1}
\end{center}
\end{figure}

\subsubsection{Phase-resolved analysis of observation 1}
We performed the same phase-resolved analysis for observation 1. As the energy resolved pulse-profiles are very similar between the observations, we used the same definition of phase bins. We also applied the same model, however, due to the overall lower statistics in observation 1 we additionally had to fix the energy of the high-energy CRSF to the phase-averaged value of 50.6\,keV. As discussed for observation 2, the energy of this CRSF does not to vary significantly as function of phase, with the possible exception during the pulse profile minimum.

We find a similar behavior of the spectral parameters as function of phase as in observation 2, albeit at much reduced statistical quality. In particular, $E_\text{CRSF,2}$ shows the same single peaked variation, with the extrema phase shifted with respect to the main peaks in the pulse profiles. 
The variation of the photon-index seems to be somewhat different, as phase-bin A now shows a relatively low $\Gamma$ instead of a high one. This might be related to the fact that in this bin the hard power-law flux is lower than that of the soft component, while the opposite is true for this bin in observation 2. This behavior might be due to a slight model intrinsic degeneracy between the photon-index $\Gamma$ and the flux ratio of the two \npex components. However, we tested for such a degeneracy in the phase-averaged spectrum by calculating a confidence contour between $\Gamma$ and the flux ratio and did not find any indications for its existence.

\section{Discussion \& Conclusion}
\label{sec:disc}

We have presented an in-depth analysis of the hard X-ray spectrum of \gx, as observed with \nustar in 2014 and 2015. We find evidence for two CRSFs, one at 35\,keV and the other at 50\,keV. Both features are broad and overlap strongly, however, thanks to \nustar's unprecedented spectral resolution above 10\,keV, the lines can be disentangled. Together they remove a significant fraction of the continuum flux.

Our results therefore confirm and reconcile the previous measurements of the CRSF. In particular the measured high values of the CRSF energy by \citet{labarbera05a} are in agreement with our high energy line, while the line detected around 35--40\,keV in, e.g., \xte and \suz \citep{kreykenbohm04a, suchy12a} corresponds to our lower energy line. 
Due to the lack of high energy coverage (e.g., in the case of \suz/PIN) or lower energy resolution (as in the \sax data), it was so far not possible to disentangle both lines in a single data set.

By applying our model to the archival \suz data, we show that those data are fully consistent with the two CRSFs proposed here. Furthermore, we find no significant change in energy as function of luminosity, despite covering about a factor of 5 in flux between \nustar observation 1 and \suz. Previous indications of a strong dependence of the line energy on luminosity are therefore very likely due to not being able to disentangle both lines.

\subsection{The location and nature of the CRSF}

Measuring two or more CRSFs in an accreting HMXBs is not unusual \citep[e.g., 4U\,0115+63 with up to 5 lines;][]{heindl99a}. In this case, the lines are interpreted as harmonics of the fundamental line, i.e., produced through scattering with electrons in higher Landau levels. The line energies should therefore be harmonically related at multiples of the fundamental energy. Here, we find that the two lines are not related at the expected factor of 2, instead we find a factor of $\sim1.4$. While  relativistic effects can cause the ratio to  deviate slightly from 2 \citep{pottschmidt05a, mueller13a}, the difference is clearly much larger than expected \citep[see, e.g.,][]{meszaros92a}. 

It is possible that the low-energy CRSF is not the fundamental line, but the first harmonic. In that case we would expect a fundamental line around 17\,keV, which gives harmonics at $\sim$34\,keV and $\sim$51\,keV, as observed.
However, the spectra contain no indication for  a feature around that energy, for any continuum model. Adding another CRSF at this energy, with a fixed width of 3\,keV results in a very small equivalent width of $\left(-70\pm110\right)$\,eV which is consistent with zero.

The fundamental line can often be much shallower than the harmonics due to photon-spawning \citep{schwarm17b} in particular in a two-column geometry, where it can almost vanish at all observation angles (Falkner et al., subm.). However, it seems unlikely that this effect would result in a perfectly filled line, without any emission wings or other deviations from a smooth spectrum.

Another possibility would be that we observe only one line, but with a significant deviation from the Gaussian shape. Such deviations are theoretically predicted, in particular in the fundamental line \citep{schoenherr07a, schwarm17b}, and have been observed in other sources, e.g., Cep~X-4 \citep{cepx4}, V\,0332+53 \citep[but see \citealt{doroshenko17a}  for further discussion]{pottschmidt05a}, and 4U\,1626$-$26 (Iwakiri et al., in prep.). However, in those cases the secondary absorption features were much weaker compared to the main line. In \gx, the lower energy feature is only about a factor four weaker than the high energy one and more easily visible as a stand-alone line.

It is also possible that both CRSFs are formed in the same accretion column, but at different locations within the column.
 If we assume that the 50\,keV line is produced at the surface of a standard neutron star with $r=10\,$km radius and the field behaves like a dipole, we can estimate the altitude at which the lower energy line is formed. This assumption implies an altitude of about 1.4\,km, taking into account the reduction of the gravitational redshift $z$ as well. 
If both lines are formed within a single column, this approach would therefore require two line forming regions: in a shock about 1.4\,km above the surface, and at the surface of the neutron star itself \citep[\textit{cf.}][]{poutanen13a}.

According to the calculations by \citet{becker12a}, \gx is in the gas shock regime, where they postulate that the CRSF is formed at the neutron star's surface. However, this is only the case for a canonical neutron star of mass $1.4\,\msun$. For a slightly more massive neutron star of $1.8\,\msun$, similar to what is postulated for Vela~X-1 \citep{quaintrell03a}, \gx shifts into the Coulomb-radiation dominated deceleration regime. In this regime, we can calculate the expected shock height via
\begin{equation}
h = 1.48\times10^5 \, L_{37}^{-5/7} B_{12}^{-4/7}\,\text{cm} 
\end{equation}
where $L_{37}$ is the luminosity in $10^{37}$\,\ergs and where we assume 10\,km for the radius of the neutron star \citep[see Eq. (51) in ][]{becker12a}.
Using the values we measure in observation 2 ($L_{37} = 0.2782\pm0.0017$, $B_{12} = 5.58^{+0.15}_{-0.14}$, corrected for the gravitational redshift at the neutron star surface for a  neutron star with a mass of 1.8\,\msun) we find $h = 1.48 \pm 0.03$\,km (uncertainties are given only by propagating the statistical uncertainties of the measurements). This value is in very good agreement with the estimate from the CRSF energy.

Note, however that in this regime we expect a positive correlation of the CRSF energy with luminosity, as the shock should be increasing in altitude with increased mass accretion rate. Such a correlation has been observed in Vela~X-1and Cep~X-4 at a similar luminosity \citep{velax1nustar, vybornov17a}. However, a correlation of the strength as observed in Cep~X-4 is within the uncertainties of our values, and we therefore cannot rule it out nor confirm its presence.

\subsection{Modeling of the phase-dependence of the CRSF energy}
\label{susec:LBmodel}

Further information about the accretion and emission geometries can be obtained from the phase dependence of the CRSF energy. 
Here we concentrate on observation 2 due to the better constraints on all spectral parameters.

The 35\,keV-line shows a very significant, almost sinusoidal variation as 
function of phase (Fig.~\ref{fig:phasresobs2}\textit{g}). This variation could indicate that we sample different 
heights of the accretion column with different intrinsic magnetic field 
strengths over the rotational phase of the neutron star. However, to have such a 
clear variation, this means that at each phase we can only observe a very small 
region of the column, which could only be realized if the emission profile is very narrow 
and altitude dependent. It is unclear how such an emission profile would physically be 
produced. 

Within the accretion column, the in-falling material can obtain relativistic 
velocities of about 0.3--0.5\,c \citep{basko76a}. This will result in the 
emitted radiation being strongly boosted towards the neutron star surface, and 
respectively experiencing a significant red-shift for the observer.  
As at different phases the angle of our line-of-sight towards the velocity vector changes, we will observe different boosting factors and therefore different energies.

In this 
picture the amplitude of the phase-dependence of the CRSF energy strongly depends on the velocity 
in the line-forming region. If the 50\,keV line is formed close to the surface, 
its velocity is basically 0, in agreement with the observed variation of energy against phase, which is compatible  with a negligible phase-dependence of that 
line. On the other hand, the 35\,keV  CRSF is formed much higher in the column, 
experiencing a strong relativistic redshift. Here we present a simple model 
based on this idea. More details for the model setup will be described in a 
forthcoming publication (Falkner et al., subm.).

The model features a single accretion column with a negligible radius extended 
only in height. The observable energy,
\begin{equation}\label{eq:E}
 E = E' \sqrt{1-\frac{R_\mathrm{S}}{R}}\quad
\frac{\sqrt{1-\beta^2}}{1+\beta \cos \eta^\star}\quad,
\end{equation}
is related to the intrinsic energy, $E'$, taking into account the gravitational 
redshift at the radius $R$ from the central mass with the Schwarzschild radius 
$R_\mathrm{S} = 2GM_\mathrm{NS}/c^2$ and the boosting due to the local bulk 
velocity $\beta=v/c$, which depends on the emission angle $\eta^\star$ with 
respect to the magnetic field (in the rest frame of the column).

In the simple picture of a cylindrical accretion column with negligible radius, 
$\eta^\star$ can be identified with the emission angle in the rest frame of the 
accretion column $\alpha$ \citep{beloborodov02a}
 \begin{equation}\label{eq:lb}
\begin{aligned}
 \cos \eta^\star & = \cos\alpha = 1 - (1-\cos\varPsi)\left(1-\frac{R_\mathrm{S}}{R}\right)\quad,
\end{aligned}
\end{equation}
accounting for general relativistic light bending based on the Schwarzschild 
metric, where $\varPsi$ is the apparent emission angle.

The apparent emission angle is related to the geometrical setup by
\begin{equation}\label{eq:angle}
 \cos\varPsi = \cos i \cos \Theta_\mathrm{AC} + \sin i \sin \Theta_\mathrm{AC} 
\cos(\Phi-\Phi_\mathrm{AC})\quad,
\end{equation}
where $i$ is the observer inclination, $\Phi$ the rotational phase, and 
$\Phi_\mathrm{AC}$ and $\Theta_\mathrm{AC}$ are the phase offset and the polar 
angle of the accretion column, respectively (Falkner et al., subm.).

The intrinsic energy $E'_\text{CRSF}(h)$ in the rest-frame of the emitting 
plasma, which is emitted at a certain height $h$ above the surface can be 
related to the observed surface energy $E_\text{CRSF}(0)$, where we 
assume $\beta=0$, by 
\begin{equation}\label{eq:Erel}
 E'_\text{CRSF}(h) = E_\text{CRSF}(0) 
\frac{R_\mathrm{NS}^3}{(R_\mathrm{NS}+h)^3 
\sqrt{1-\frac{R_\mathrm{S}}{R_\mathrm{ NS}}}}.
\end{equation}
Equation~\eqref{eq:Erel} accounts for the gravitational redshift and the 
decrease of the dipolar magnetic field. Further $E_\text{CRSF}(0)$ is a direct 
tracer of the surface magnetic field. Here we assume that the observed surface 
energy $E_\text{CRSF}(0)$ is the phase-averaged observed energy of the 
high-energy CRSF line, 49.6\,keV, and therefore that this line does not vary significantly as function of phase.

\begin{figure*}
 \includegraphics[width=0.9\textwidth]{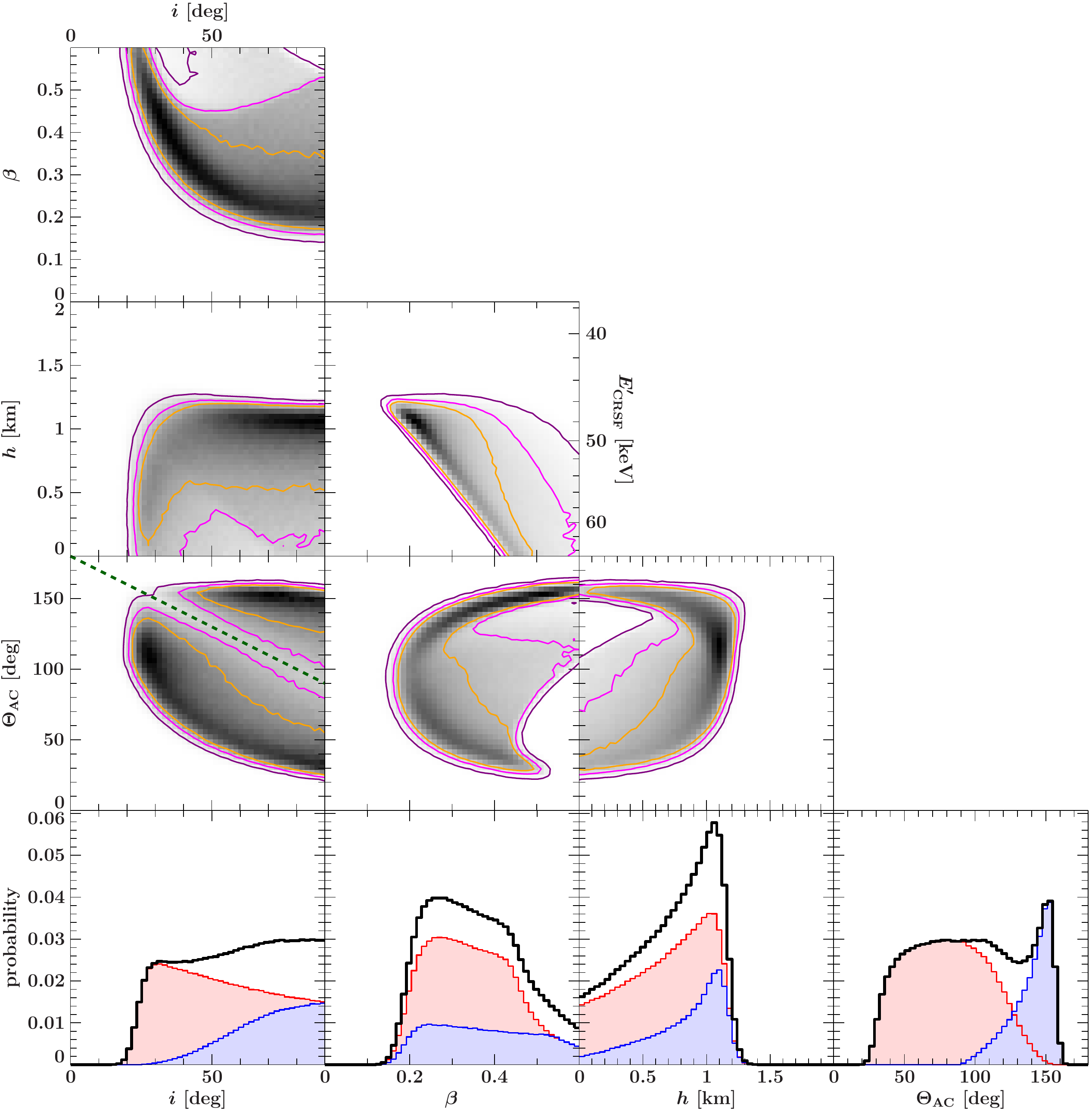}
 \caption{Parameter probabilities for the model described in Sect.~\ref{susec:LBmodel} obtained from Monte Carlo simulations
(\texttt{EMCEE}). Parameter correlations are shown as color-maps, where
black
corresponds to highest probability. Purple, magenta and orange
correspond to
the 99\%, 90\% and 68\% probability level. In the $i$-$\Theta_\mathrm{AC}$
space there are two solutions visible, which are divided by
$\Theta_\mathrm{AC}=-i+180^\circ$ (green dashed line), where $i$ is the inclination and $\Theta_\mathrm{AC}$ is the polar angle of the accretion column. The bottom panels show the one dimensional parameter probabilities, where red and blue histograms
correspond to the individual solutions.}
 \label{fig:emcee}
\end{figure*}

Due to the many free parameters and degeneracies in this simple model, we 
perform MCMC calculations to estimate the parameter space using the 
\texttt{emcee} routine in ISIS. This routine is based upon the parallel ``simple
 stretch'' method presented by \citet{foreman12a}.

Figure~\ref{fig:emcee} shows the probability 
distribution of the parameters and Tab.~\ref{tab:mcmcpars} gives the resulting 
parameter constraints. We used  $50$ free walkers for each free parameter and 
$100000$ iterations. The underlying model assumes that the two CRSFs originate from 
the same accretion column. Furthermore we fix $\Phi_\mathrm{AC}$ to 0.562, 
which was found to be well constrained in a preliminary analysis and does not show any correlations to other parameters.

\begin{table}
\centering
  \caption{Parameter constraints from the MCMC runs.}
  \label{tab:mcmcpars}
 \begin{tabular}[b]{l c r }
  \hline\hline
  {model parameter }		& symbol 		& best-fit-values  \\
    \hline
    \textit{variable parameters} & & \\
  observer inclination	& $i$			& $>19^\circ$ \\
  local bulk velocity	& $\beta$		& $0.38^{+0.23}_{-0.11}$  \\
  height of emission	& $h$			& $<1.2$\,km\\
  polar angle of B-field& $\Theta_\mathrm{AC}$	& $22^\circ$--$160^\circ$\\\hline
  \textit{fixed parameters} && \\
  phase offset		& $\Phi_\mathrm{AC}$	& $0.562^\circ\pm0.025$  \\
  observed CRSF energy at $h=0$ & $E(0)$	& $49.6$\, keV\\
  neutron star mass	& $M_\mathrm{NS}$	& $1.4\,M_\odot$\\
  neutron star radius	& $R_\mathrm{NS}$	& $10$\,km\\
  \hline
  \end{tabular}
\end{table}

\begin{figure}
\begin{center}
\includegraphics[width=0.95\columnwidth]{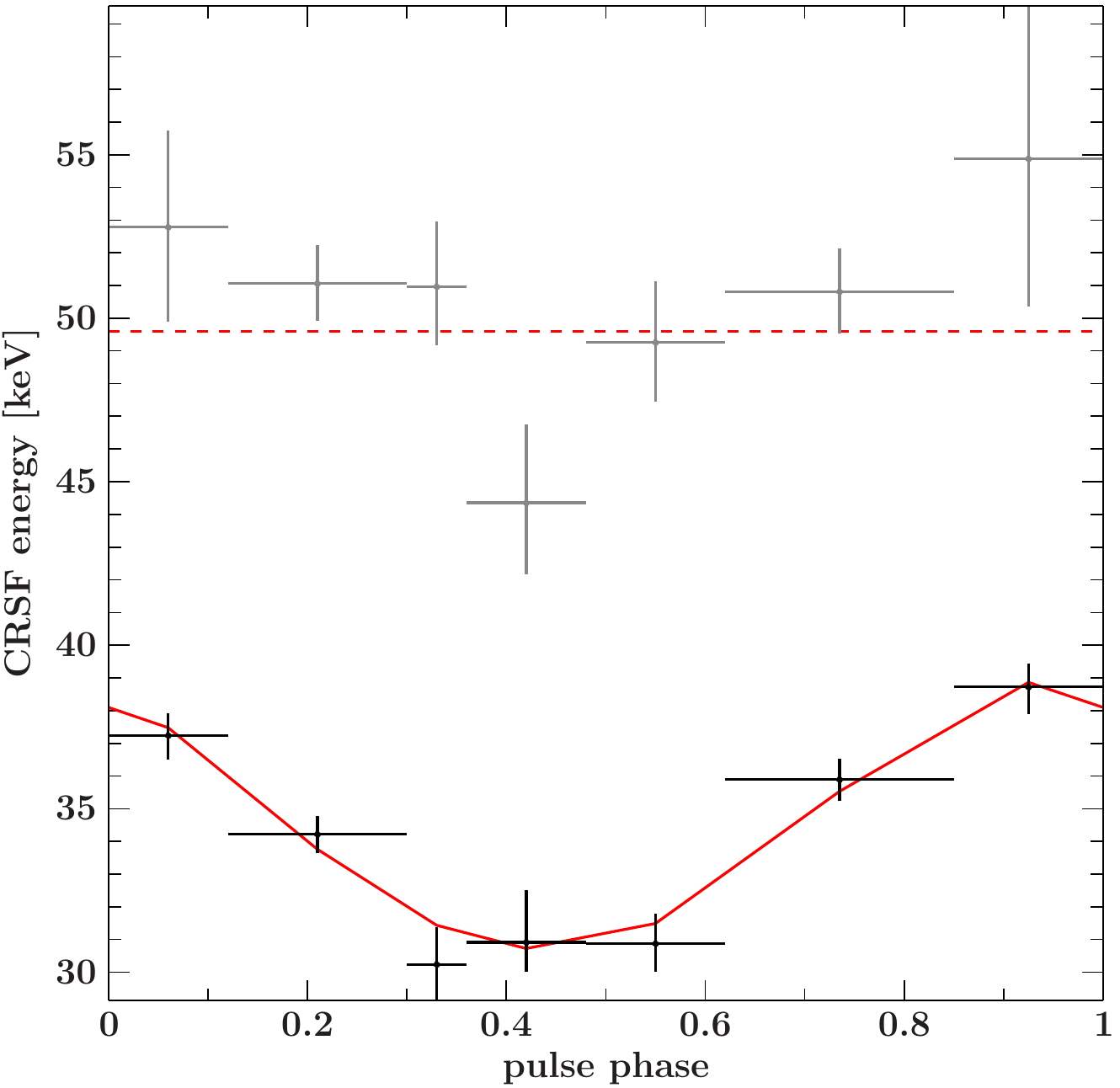}
\caption{Phase-dependence of the CRSF energies during \nustar observation 2, together with the predicted energies by our simple accretion column model in red. The energy of the higher energy line as been fixed to 49.6\,keV in the model.}
\label{fig:crsfenergy}
\end{center}
\end{figure}

This simple model provides an excellent description of the observed 
phase-dependence of the CRSF energy, as shown in Figure~\ref{fig:crsfenergy}. It 
also provides some limits on the geometry of the system, for example it 
indicates an inclination $i>20^\circ$ and a velocity $\beta$ between 0.2--0.4\,$c$. These velocities are well in agreement with theoretical calculations.
We also find a column height of around 
1\,km, which is very similar to the height we estimated independently from the 
shock height in the accretion column model by \citet{becker12a}.

We note that in a further step we could assume a certain velocity profile within the accretion column, e.g., the shock model as discussed by \citet{becker12a}. In this case, the velocity below the shock would be much slower, and the emission region must be constrained to be around 1\,km. This will lead to  better constraints of the geometric parameters. However, a detailed calculation of this model is beyond the scope of the this paper.

Of course, a possible solution could also include contributions from both 
accretion columns. However, we do not consider this case here, as we cannot 
constrain the relative contribution of each column to the observed flux with the 
available data. In such a model, we would need to make sophisticated 
assumptions about the altitude-dependent emission profile in each column, which 
is not necessary in the present set-up.

\subsection{Summary}
The very high data quality of the \nustar observations of \gx has allowed us to identify two CRSFs, one at 34\,keV and one at 51\,keV, where the secondary line feature is statistically significant at $\ge99.9\%$. At the same time, we rule out with high confidence a fundamental line at 17\,keV, and therefore the possibility that the two observed lines are harmonically related. We speculate on different origins for the two lines, but current available theoretical models do not yet allow to draw firm conclusions. 
We argue that a likely scenario is a symmetric magnetic field (and accretion) geometry, in which the CRSFs are formed at two different altitudes above the neutron star: the low-energy line at about 1--1.4\,km above the surface in a shock, and the high-energy line at the surface of the star. Both lines are therefore interpreted as the fundamental line corresponding to the magnetic field strength at the respective line forming altitude.

Based on this model, we calculated the expected energy variance of the lines, solely due to the changing viewing angle with phase and the boosting in the relativistic velocities of the in-falling material. Here we do not make any assumptions about the configuration and emission profile of the accretion column. This simple model provides a very good description of the phase-dependence of the CRSF energy and gives some constraints on the geometry of the neutron star and its magnetic field. Further observations of the phase-dependence at different luminosities and theoretical calculations are needed to confirm this model.

\hspace{\baselineskip}

\begin{acknowledgements}
We thank the referee for their helpful comments.
This research has made use of the {\it NuSTAR}
Data Analysis Software (NuSTARDAS) jointly developed by the ASI
Science Data Center (ASDC, Italy) and the California Institute of
Technology (USA). 
We would like to thank John E. Davis for the \texttt{slxfig} module, which was used to produce all figures in this work.
The \textsl{Swift}/BAT transient monitor results were
provided by the \textsl{Swift}/BAT team.
This research has made use of data obtained from the \textsl{Suzaku} satellite, a collaborative mission between the space agencies of Japan (JAXA) and the USA (NASA).
L.N. acknowledges support by ASI/INAF grant I/037/12/0 and PRIN-INAF
2014 grant ``Towards a unified picture of accretion in High Mass X-Ray
Binaries''.

\end{acknowledgements}

\begin{appendix}
\section{Fit parameters for all phase-averaged models}
\label{sec:allfits}
Tables~\ref{tab:obs2all_gabs1} and \ref{tab:obs2all_gabs2} give the best-fit parameters for all tested continuum models, using one and two CRSFs, respectively. 
The values for the \texttt{NPEX}, \texttt{FDcut}, and \texttt{HIGHECUT} model given Table~\ref{tab:obs2all_gabs2} are the same as in Table~\ref{tab:phasavg}, but repeated for easy reference.
Note that in particular the \texttt{cutoffpl} model did not lead to an acceptable fit, and the given uncertainties on the fit parameters are therefore likely underestimated and should be regarded with care. In this model we also had to allow for a secondary, very high absortion column (\nhtwo) as well as fix the energy of the \fekb line to 7.05\,keV.

\begin{sidewaystable*}
\caption{Best-fit model parameters for the phase-averaged observation 1 spectra using only one GABS component to model the CRSF.}\label{tab:obs1all_gabs1}
\centering
\begin{tabular}{r|lllll}
\hline\hline
 Parameter & CPL & NPEX & FDcut & HighECut & COMPMAG  \\\hline
 $ N_\text{H,1}~(10^{22}\,\text{cm}^{-2})$ & $21.7\pm0.7$ & $39.9\pm1.5$ & $37.8^{+1.5}_{-1.6}$ & $40.0\pm1.3$ & $39.0\pm1.5$ \\
 $ N_\text{H,2}~(10^{22}\,\text{cm}^{-2})$ & $\left(4.00^{+0.00}_{-0.04}\right)\times10^{2}$ & --- & --- & --- & --- \\
 $ f$ & $0.656\pm0.016$ & $0.913\pm0.005$ & $0.920\pm0.005$ & $0.919^{+0.005}_{-0.004}$ & $0.897\pm0.006$ \\
 $ \mathcal{F}~(10^{-9}\,\text{erg}\,\text{cm}^{-2}\,\text{s}^{-1})$\tablefootmark{a}$$ &  $2.06\pm0.04$ & $1.757\pm0.012$ & $1.745\pm0.012$ & $1.762\pm0.011$ & $1.746\pm0.012$ \\
 $ A_\text{2}$\tablefootmark{b}$$ & --- & $\left(5.9^{+0.8}_{-0.4}\right)\times10^{-5}$ & --- & --- & --- \\
 $ \Gamma$ & $0.65\pm0.04$ & $0.85\pm0.04$ & $1.409^{+0.024}_{-0.027}$ & $1.455^{+0.023}_{-0.028}$ & --- \\
 $ E_\text{cut}~(\text{keV})$ & --- & --- & $42\pm5$ & $23.8^{+1.0}_{-0.9}$ & --- \\
 $ E_\text{fold}~(\text{keV})$ & $12.88^{+0.30}_{-0.29}$ & $6.90^{+0.09}_{-0.33}$ & $5.4^{+1.0}_{-0.9}$ & $14.7^{+1.0}_{-1.9}$ & --- \\
 $ kT_\text{BB}$ & --- & --- & --- & --- & $6.34^{+0.40}_{-0.06}$ \\
 $ kT_\text{e}$ & --- & --- & --- & --- & $5.1^{+3.2}_{-0.8}$ \\
 $ \tau$ & --- & --- & --- & --- & $0.0105^{+0.0520}_{-0.0006}$ \\
 $ A_\text{disk}$ & --- & --- & --- & --- & $8.7^{+1.4}_{-7.5}$ \\
 $ kT_\text{disk}$ & --- & --- & --- & --- & $2.67\pm0.08$ \\
 $ \text{EW}_{\text{K}\alpha}~(\text{eV})$ & $\left(1.90\pm0.07\right)\times10^{2}$ & $\left(1.08^{+0.08}_{-0.07}\right)\times10^{2}$ & $\left(1.22^{+0.11}_{-0.10}\right)\times10^{2}$ & $\left(1.09\pm0.07\right)\times10^{2}$ & $\left(1.07^{+0.08}_{-0.07}\right)\times10^{2}$ \\
 $ E_{\text{K}\alpha}~(\text{keV})$ & $6.325\pm0.014$ & $6.3547^{+0.0015}_{-0.0334}$ & $6.339^{+0.012}_{-0.013}$ & $6.3548^{+0.0012}_{-0.0335}$ & $6.334^{+0.022}_{-0.013}$ \\
 $ \sigma_{\text{K}\alpha}~(\text{keV})$ & $0.171^{+0.029}_{-0.028}$ & $\le0.06$ & $0.05^{+0.04}_{-0.06}$ & $\le0.06$ & $\le0.06$ \\
 $ E_{\text{K}\beta}~(\text{keV})$ & --- & --- & --- & --- & $7.05^{+0.15}_{-0.16}$ \\
 $ \text{EW}_{\text{K}\beta}~(\text{eV})$ & $8\pm5$ & $-5^{+6}_{-5}$ & $-0\pm6$ & $-5\pm5$ & $-9^{+6}_{-5}$ \\
 $ E_\text{CRSF,1}~(\text{keV})$ & $51.3\pm0.6$ & $49.5^{+0.6}_{-1.8}$ & $42.1\pm2.2$ & $54.1^{+1.8}_{-3.7}$ & $48.1^{+2.4}_{-2.9}$ \\
 $ \sigma_\text{CRSF,1}~(\text{keV})$ & $10.6\pm0.4$ & $12.9^{+0.5}_{-1.5}$ & $9.2^{+0.8}_{-1.2}$ & $14.3^{+1.6}_{-3.2}$ & $13.3^{+2.2}_{-2.7}$ \\
 $ d_\text{CRSF,1}~(\text{keV})$ & --- & $40^{+0}_{-12}$ & $29^{+12}_{-13}$ & $30^{+0}_{-11}$ & $35^{+16}_{-17}$ \\
 $ CC_\text{B}$ & $1.040\pm0.004$ & $1.040\pm0.004$ & $1.040\pm0.004$ & $1.040\pm0.004$ & $1.040\pm0.004$ \\
 $ \text{GS~(eV)}$ & $0.031\pm0.011$ & $0.037\pm0.009$ & $0.034\pm0.010$ & $0.035\pm0.009$ & $0.035\pm0.009$ \\
$\mathcal{L}~(10^{36}$\,erg\,s$^{-1}$)\tablefootmark{c} & $1.132^{+0.028}_{-0.027}$ & $1.575^{+0.009}_{-0.008}$ & $1.586^{+0.009}_{-0.008}$ & $1.586\pm0.007$ & $1.547^{+0.010}_{-0.009}$ \\
\hline$\chi^2/\text{d.o.f.}$   & 1020.68/458& 607.41/457& 650.74/457& 566.60/455& 600.77/454\\$\chi^2_\text{red}$   & 2.229& 1.329& 1.424& 1.245& 1.323\\\hline
\end{tabular}
\tablefoot{\tablefoottext{a}{unabsorbed  flux between 5--50\,keV}
\tablefoottext{b}{in ph\,keV$^{-1}$\,cm$^{-2}$\,s$^{-1}$~at 1\,keV}
\tablefoottext{c}{luminosity between 5--50\,keV for a distance of 3.0\,kpc}
}
\end{sidewaystable*}

\begin{sidewaystable*}
\caption{Best-fit model parameters for the phase-averaged observation 1 spectra using two GABS components for the CRSF (see also Tab.~
ef{tab:phasavg}).}\label{tab:obs1all_gabs2}
\centering
\begin{tabular}{r|lllll}
\hline\hline
 Parameter & CPL & NPEX & FDcut & HighECut & COMPMAG  \\\hline
 $ N_\text{H,1}~(10^{22}\,\text{cm}^{-2})$ & $21.6^{+0.7}_{-0.8}$ & $39.8\pm1.5$ & $39.6\pm1.3$ & $39.9^{+1.3}_{-1.4}$ & $39.3\pm1.5$ \\
 $ N_\text{H,2}~(10^{22}\,\text{cm}^{-2})$ & $\left(5.0^{+0.5}_{-0.4}\right)\times10^{2}$ & --- & --- & --- & --- \\
 $ f$ & $0.666\pm0.022$ & $0.910\pm0.006$ & $0.9201^{+0.0010}_{-0.0041}$ & $0.921^{+0.005}_{-0.004}$ & $0.899\pm0.006$ \\
 $ \mathcal{F}~(10^{-9}\,\text{erg}\,\text{cm}^{-2}\,\text{s}^{-1})$\tablefootmark{a}$$ & $2.08\pm0.05$ & $1.757\pm0.012$ & $1.760\pm0.011$ & $1.762\pm0.011$ & $1.750\pm0.012$ \\
 $ A_\text{2}$\tablefootmark{b}$$ & --- & $\left(6.8^{+0.9}_{-0.8}\right)\times10^{-5}$ & --- & --- & --- \\
 $ \Gamma$ & $0.62\pm0.05$ & $0.788\pm0.030$ & $1.454^{+0.021}_{-0.025}$ & $1.467^{+0.020}_{-0.022}$ & --- \\
 $ E_\text{cut}~(\text{keV})$ & --- & --- & $45.5^{+4.1}_{-2.2}$ & $37.3^{+0.8}_{-1.0}$ & --- \\
 $ E_\text{fold}~(\text{keV})$ & $12.8^{+0.6}_{-0.5}$ & $6.52^{+0.35}_{-0.23}$ & $6.3^{+1.0}_{-1.1}$ & $8.89^{+0.85}_{-0.20}$ & --- \\
 $ kT_\text{BB}$ & --- & --- & --- & --- & $5.32\pm0.22$ \\
 $ kT_\text{e}$ & --- & --- & --- & --- & $9.4^{+0.7}_{-9.2}$ \\
 $ \tau$ & --- & --- & --- & --- & $0.062^{+0.040}_{-0.021}$ \\
 $ A_\text{disk}$ & --- & --- & --- & --- & $0.62^{+0.52}_{-0.13}$ \\
 $ kT_\text{disk}$ & --- & --- & --- & --- & $2.48^{+0.10}_{-0.09}$ \\
 $ \text{EW}_{\text{K}\alpha}~(\text{eV})$ & $\left(1.87^{+0.16}_{-0.15}\right)\times10^{2}$ & $\left(1.08^{+0.08}_{-0.07}\right)\times10^{2}$ & $\left(1.49^{+0.09}_{-0.08}\right)\times10^{2}$ & $\left(1.10\pm0.07\right)\times10^{2}$ & $\left(1.07\pm0.09\right)\times10^{2}$ \\
 $ E_{\text{K}\alpha}~(\text{keV})$ & $6.328\pm0.014$ & $6.352^{+0.005}_{-0.032}$ & $6.355795^{+0.000011}_{-0.034438}$ & $6.3548^{+0.0011}_{-0.0335}$ & $6.334^{+0.022}_{-0.013}$ \\
 $ \sigma_{\text{K}\alpha}~(\text{keV})$ & $0.167^{+0.029}_{-0.028}$ & $\left(1.6^{+49.9}_{-1.6}\right)\times10^{-3}$ & $\le0.07$ & $\le0.06$ & $\le0.06$ \\
 $ E_{\text{K}\beta}~(\text{keV})$ & --- & --- & --- & --- & $7.05^{+0.15}_{-0.16}$ \\
 $ \text{EW}_{\text{K}\beta}~(\text{eV})$ & $6\pm6$ & $-6\pm6$ & $10\pm5$ & $-4^{+6}_{-5}$ & $-9\pm6$ \\
 $ E_\text{CRSF,1}~(\text{keV})$ & $54.2^{+1.2}_{-1.0}$ & $50.6^{+2.1}_{-1.7}$ & $50.2714^{+0.0007}_{-1.7640}$ & $48.9^{+3.4}_{-2.2}$ & $50.9^{+2.7}_{-2.3}$ \\
 $ \sigma_\text{CRSF,1}~(\text{keV})$ & $9.6\pm0.7$ & $8.8^{+1.2}_{-2.3}$ & $8.0090^{+0.1674}_{-0.0011}$ & $9.0^{+1.4}_{-2.3}$ & $8.8^{+1.2}_{-2.6}$ \\
 $ d_\text{CRSF,1}~(\text{keV})$ & --- & $20^{+9}_{-7}$ & --- & $27.6^{+2.5}_{-8.9}$ & $18^{+6}_{-7}$ \\
 $ E_\text{CRSF,2}~(\text{keV})$ & $36.0^{+1.4}_{-1.1}$ & $34.7^{+2.1}_{-1.4}$ & $35.7^{+2.2}_{-1.7}$ & $31.3^{+6.8}_{-1.3}$ & $34.7^{+1.9}_{-1.2}$ \\
 $ \sigma_\text{CRSF,2}~(\text{keV})$ & $4.5^{+0.9}_{-0.8}$ & $5.0^{+1.1}_{-1.2}$ & $6.3199^{+0.8459}_{-0.0023}$ & $5.0^{+2.1}_{-0.9}$ & $4.3\pm1.1$ \\
 $ d_\text{CRSF,2}~(\text{keV})$ & $3.5^{+1.9}_{-1.3}$ & $3.8^{+3.3}_{-2.1}$ & $11.2906^{+7.5606}_{-0.0011}$ & $4.6^{+20.4}_{-2.0}$ & $2.6^{+2.1}_{-1.3}$ \\
 $ CC_\text{B}$ & $1.040\pm0.004$ & $1.040\pm0.004$ & $1.040\pm0.004$ & $1.040\pm0.004$ & $1.040\pm0.004$ \\
 $ \text{GS~(eV)}$ & $0.029\pm0.011$ & $0.036^{+0.009}_{-0.010}$ & $0.035\pm0.009$ & $0.036\pm0.009$ & $0.035\pm0.009$ \\
$\mathcal{L}~(10^{36}$\,erg\,s$^{-1}$)\tablefootmark{c} & $1.15\pm0.04$ & $1.569\pm0.009$ & $1.5874^{+0.0018}_{-0.0070}$ & $1.588\pm0.007$ & $1.552^{+0.010}_{-0.009}$ \\
\hline$\chi^2/\text{d.o.f.}$   & 863.94/455& 552.65/454& 556.28/455& 549.93/452& 561.16/451\\$\chi^2_\text{red}$   & 1.899& 1.217& 1.223& 1.217& 1.244\\\hline
\end{tabular}
\tablefoot{\tablefoottext{a}{unabsorbed  flux between 5--50\,keV}
\tablefoottext{b}{in ph\,keV$^{-1}$\,cm$^{-2}$\,s$^{-1}$~at 1\,keV}
\tablefoottext{c}{luminosity between 5--50\,keV for a distance of 3.0\,kpc}
}
\end{sidewaystable*}

\begin{sidewaystable*}
\caption{Best-fit model parameters for the phase-averaged fits of observation 2 using only one GABS component to model the CRSF.}\label{tab:obs2all_gabs1}
\centering
\begin{tabular}{r|lllll}
\hline\hline
 Parameter & CPL & NPEX & FDcut & HighECut & COMPMAG  \\\hline
 $ N_\text{H,1}~(10^{22}\,\text{cm}^{-2})$ & $12.9\pm0.5$ & $29.7\pm1.6$ & $23.9\pm1.7$ & $26.4^{+1.6}_{-1.7}$ & $29.78^{+0.93}_{-0.10}$ \\
 $ N_\text{H,2}~(10^{22}\,\text{cm}^{-2})$ & $\left(4.83^{+0.23}_{-0.21}\right)\times10^{2}$ & --- & --- & --- & --- \\
 $ f$ & $0.548\pm0.014$ & $0.832\pm0.009$ & $0.834^{+0.016}_{-0.013}$ & $0.829^{+0.012}_{-0.010}$ & $0.7956^{+0.0040}_{-0.0024}$ \\
 $ \mathcal{F}~(10^{-9}\,\text{erg}\,\text{cm}^{-2}\,\text{s}^{-1})$\tablefootmark{a}$$ & $3.47\pm0.07$ & $2.581\pm0.016$ & $2.532\pm0.015$ & $2.551\pm0.015$ & $2.569^{+0.004}_{-0.008}$ \\
 $ A_\text{2}$\tablefootmark{b}$$ & --- & $\left(1.22^{+0.15}_{-0.11}\right)\times10^{-4}$ & --- & --- & --- \\
 $ \Gamma$ & $0.649^{+0.030}_{-0.031}$ & $0.90\pm0.04$ & $1.270^{+0.022}_{-0.023}$ & $1.327^{+0.020}_{-0.021}$ & --- \\
 $ E_\text{cut}~(\text{keV})$ & --- & --- & $36.7^{+3.1}_{-1.8}$ & $22.3^{+0.5}_{-0.4}$ & --- \\
 $ E_\text{fold}~(\text{keV})$ & $12.6\pm0.4$ & $6.58^{+0.16}_{-0.30}$ & $4.9^{+0.4}_{-0.6}$ & $11.3^{+0.8}_{-0.7}$ & --- \\
 $ kT_\text{BB}$ & --- & --- & --- & --- & $5.985^{+0.021}_{-0.004}$ \\
 $ kT_\text{e}$ & --- & --- & --- & --- & $10^{+0}_{-8}$ \\
 $ \tau$ & --- & --- & --- & --- & $\le0.03$ \\
 $ A_\text{disk}$ & --- & --- & --- & --- & $2.79\pm0.08$ \\
 $ kT_\text{disk}$ & --- & --- & --- & --- & $2.4420^{+0.0028}_{-0.0037}$ \\
 $ \text{EW}_{\text{K}\alpha}~(\text{eV})$ & $\left(1.74\pm0.09\right)\times10^{2}$ & $\left(1.35\pm0.08\right)\times10^{2}$ & $\left(1.61^{+0.10}_{-0.09}\right)\times10^{2}$ & $\left(1.47\pm0.06\right)\times10^{2}$ & --- \\
 $ E_{\text{K}\alpha}~(\text{keV})$ & $6.332\pm0.009$ & $6.336^{+0.021}_{-0.009}$ & $6.332\pm0.009$ & $6.333\pm0.009$ & $6.334^{+0.015}_{-0.009}$ \\
 $ \sigma_{\text{K}\alpha}~(\text{keV})$ & $0.102^{+0.018}_{-0.019}$ & $0.024^{+0.032}_{-0.024}$ & $0.082^{+0.020}_{-0.023}$ & $0.058^{+0.023}_{-0.033}$ & $0.013^{+0.031}_{-0.013}$ \\
 $ E_{\text{K}\beta}~(\text{keV})$ & --- & $7.05^{+0.45}_{-0.10}$ & $7.035^{+0.012}_{-0.155}$ & $7.04^{+0.06}_{-0.16}$ & $7.05^{+0.15}_{-0.16}$ \\
 $ \text{EW}_{\text{K}\beta}~(\text{eV})$ & $8\pm5$ & $7\pm5$ & $10\pm5$ & $7\pm4$ & --- \\
 $ E_\text{CRSF,1}~(\text{keV})$ & $52.5\pm0.5$ & $48.7^{+0.9}_{-1.6}$ & $37.7^{+1.9}_{-1.2}$ & $49.2\pm1.7$ & $47.0^{+0.5}_{-1.5}$ \\
 $ \sigma_\text{CRSF,1}~(\text{keV})$ & $12.7\pm0.4$ & $12.8^{+0.7}_{-1.4}$ & $7.6^{+0.9}_{-0.7}$ & $9.6\pm1.3$ & $12.19^{+0.13}_{-1.35}$ \\
 $ d_\text{CRSF,1}~(\text{keV})$ & $50.0^{+0.0}_{-2.8}$ & $46^{+5}_{-12}$ & $18^{+10}_{-5}$ & $14^{+6}_{-5}$ & $34.0^{+0.4}_{-1.4}$ \\
 $ CC_\text{B}$ & $1.0373\pm0.0030$ & $1.0375\pm0.0030$ & $1.0372\pm0.0030$ & $1.0373\pm0.0030$ & $1.0374^{+0.0022}_{-0.0028}$ \\
 $ \text{GS~(eV)}$ & $0.018\pm0.009$ & $0.024\pm0.009$ & $0.017\pm0.009$ & $0.020\pm0.009$ & $0.024^{+0.009}_{-0.007}$ \\
$\mathcal{L}~(10^{36}$\,erg\,s$^{-1}$)\tablefootmark{c} & $0.946^{+0.023}_{-0.024}$ & $1.436\pm0.015$ & $1.438^{+0.027}_{-0.022}$ & $1.431^{+0.020}_{-0.017}$ & $1.373^{+0.007}_{-0.005}$ \\
\hline$\chi^2/\text{d.o.f.}$   & 878.91/457& 542.81/456& 744.03/456& 564.24/454& 523.52/454\\$\chi^2_\text{red}$   & 1.923& 1.190& 1.632& 1.243& 1.153\\\hline
\end{tabular}
\tablefoot{\tablefoottext{a}{unabsorbed  flux between 5--50\,keV}
\tablefoottext{b}{in ph\,keV$^{-1}$\,cm$^{-2}$\,s$^{-1}$~at 1\,keV}
\tablefoottext{c}{luminosity between 5--50\,keV for a distance of 3.0\,kpc}
}
\end{sidewaystable*}

\begin{sidewaystable*}
\caption{Best-fit model parameters for the phase-averaged fits using two GABS components for the CRSF (see also Tab.~
ef{tab:phasavg}).}\label{tab:obs2all_gabs2}
\centering
\begin{tabular}{r|lllll}
\hline\hline
 Parameter & CPL & NPEX & FDcut & HighECut & COMPMAG  \\\hline
 $ N_\text{H,1}~(10^{22}\,\text{cm}^{-2})$ & $13.4\pm0.5$ & $29.8\pm1.6$ & $25.7\pm1.6$ & $26.3\pm1.6$ & $29.8^{+1.5}_{-1.7}$ \\
 $ N_\text{H,2}~(10^{22}\,\text{cm}^{-2})$ & $\left(4.03^{+0.24}_{-0.23}\right)\times10^{2}$ & --- & --- & --- & --- \\
 $ f$ & $0.612\pm0.016$ & $0.828\pm0.009$ & $0.829^{+0.012}_{-0.011}$ & $0.830^{+0.012}_{-0.010}$ & $0.801^{+0.010}_{-0.009}$ \\
 $ \mathcal{F}~(10^{-9}\,\text{erg}\,\text{cm}^{-2}\,\text{s}^{-1})$\tablefootmark{a}$$ & $3.21\pm0.06$ & $2.583\pm0.016$ & $2.545\pm0.014$ & $2.550^{+0.015}_{-0.014}$ & $2.572^{+0.015}_{-0.016}$ \\
 $ A_\text{2}$\tablefootmark{b}$$ & --- & $\left(1.47\pm0.12\right)\times10^{-4}$ & --- & --- & --- \\
 $ \Gamma$ & $0.74\pm0.04$ & $0.840\pm0.029$ & $1.308^{+0.020}_{-0.021}$ & $1.325\pm0.020$ & --- \\
 $ E_\text{cut}~(\text{keV})$ & --- & --- & $44.2\pm2.0$ & $21.87^{+0.38}_{-0.29}$ & --- \\
 $ E_\text{fold}~(\text{keV})$ & $14.1\pm0.6$ & $6.10^{+0.17}_{-0.13}$ & $5.3^{+0.6}_{-0.7}$ & $11.8^{+1.0}_{-0.6}$ & --- \\
 $ kT_\text{BB}$ & --- & --- & --- & --- & $5.29^{+0.15}_{-0.13}$ \\
 $ kT_\text{e}$ & --- & --- & --- & --- & $10^{+0}_{-9}$ \\
 $ \tau$ & --- & --- & --- & --- & $0.043\pm0.014$ \\
 $ A_\text{disk}$ & --- & --- & --- & --- & $0.92^{+0.60}_{-0.14}$ \\
 $ kT_\text{disk}$ & --- & --- & --- & --- & $2.30^{+0.07}_{-0.06}$ \\
 $ \text{EW}_{\text{K}\alpha}~(\text{eV})$ & --- & $\left(1.34\pm0.06\right)\times10^{2}$ & $\left(1.49^{+0.09}_{-0.08}\right)\times10^{2}$ & $\left(1.46^{+0.09}_{-0.08}\right)\times10^{2}$ & $\left(1.33^{+0.09}_{-0.07}\right)\times10^{2}$ \\
 $ E_{\text{K}\alpha}~(\text{keV})$ & $6.332\pm0.009$ & $6.337^{+0.019}_{-0.008}$ & $6.333\pm0.009$ & $6.333\pm0.009$ & $6.335^{+0.020}_{-0.014}$ \\
 $ \sigma_{\text{K}\alpha}~(\text{keV})$ & $0.099^{+0.018}_{-0.019}$ & $0.022^{+0.033}_{-0.022}$ & $0.066^{+0.022}_{-0.028}$ & $0.060^{+0.023}_{-0.031}$ & $0.017^{+0.036}_{-0.018}$ \\
 $ E_{\text{K}\beta}~(\text{keV})$ & --- & $7.05^{+0.20}_{-0.10}$ & $7.04^{+0.05}_{-0.16}$ & $7.04^{+0.06}_{-0.16}$ & $7.05^{+0.15}_{-0.16}$ \\
 $ \text{EW}_{\text{K}\beta}~(\text{eV})$ & --- & $10\pm5$ & $10\pm5$ & $10\pm5$ & $4\pm5$ \\
 $ E_\text{CRSF,1}~(\text{keV})$ & $55.4^{+0.9}_{-0.8}$ & $49.6^{+1.3}_{-1.2}$ & $49.2\pm1.2$ & $50.4^{+1.8}_{-1.4}$ & $49.3^{+1.4}_{-1.2}$ \\
 $ \sigma_\text{CRSF,1}~(\text{keV})$ & $11.1\pm0.5$ & $7.8^{+2.3}_{-1.5}$ & $7.0^{+0.7}_{-0.6}$ & $7.4^{+2.1}_{-1.9}$ & $7.3^{+2.0}_{-1.8}$ \\
 $ d_\text{CRSF,1}~(\text{keV})$ & $50.0^{+0.0}_{-2.0}$ & $20^{+9}_{-5}$ & --- & $13^{+8}_{-5}$ & $16^{+7}_{-5}$ \\
 $ E_\text{CRSF,2}~(\text{keV})$ & $35.4^{+1.0}_{-0.8}$ & $34.5^{+1.6}_{-1.4}$ & $35.6\pm1.3$ & $35.1^{+1.6}_{-1.2}$ & $34.5^{+1.7}_{-1.2}$ \\
 $ \sigma_\text{CRSF,2}~(\text{keV})$ & $4.8^{+0.7}_{-0.6}$ & $5.1\pm0.8$ & $6.5\pm0.5$ & $3.5^{+1.0}_{-0.9}$ & $4.6^{+0.8}_{-0.7}$ \\
 $ d_\text{CRSF,2}~(\text{keV})$ & $4.2^{+1.7}_{-1.2}$ & $5.1^{+2.6}_{-2.5}$ & $16\pm5$ & $1.5^{+1.1}_{-0.7}$ & $3.9^{+2.3}_{-1.7}$ \\
 $ CC_\text{B}$ & $1.0372\pm0.0030$ & $1.0373\pm0.0030$ & $1.0372\pm0.0030$ & $1.0373\pm0.0030$ & $1.0374^{+0.0031}_{-0.0030}$ \\
 $ \text{GS~(eV)}$ & $0.018\pm0.009$ & $0.024\pm0.009$ & $0.019\pm0.009$ & $0.020\pm0.009$ & $0.024\pm0.009$ \\
$\mathcal{L}~(10^{36}$\,erg\,s$^{-1}$)\tablefootmark{c} & $1.056\pm0.027$ & $1.428^{+0.016}_{-0.015}$ & $1.431^{+0.021}_{-0.018}$ & $1.432^{+0.020}_{-0.017}$ & $1.381^{+0.017}_{-0.016}$ \\
\hline$\chi^2/\text{d.o.f.}$   & 717.13/454& 456.91/454& 559.94/454& 533.75/451& 460.24/451\\$\chi^2_\text{red}$   & 1.580& 1.006& 1.233& 1.183& 1.020\\\hline
\end{tabular}
\tablefoot{\tablefoottext{a}{unabsorbed  flux between 5--50\,keV}
\tablefoottext{b}{in ph\,keV$^{-1}$\,cm$^{-2}$\,s$^{-1}$~at 1\,keV}
\tablefoottext{c}{luminosity between 5--50\,keV for a distance of 3.0\,kpc}
}
\end{sidewaystable*}

\end{appendix}

\end{document}